\documentclass[sigconf]{acmart}

\usepackage{epsfig,endnotes}
\usepackage{type1cm}	
\usepackage{graphicx}     
\usepackage{xspace}     
\usepackage{balance}     
\usepackage{booktabs}     
\usepackage[tableposition=top,font={small}]{caption}     
\usepackage{siunitx}	
\usepackage{epsfig}
\usepackage{framed}
\usepackage{newfloat}
\usepackage{amsmath, amssymb, eurosym}
\usepackage{enumerate}
\usepackage{enumitem}
\usepackage[vlined,linesnumbered,ruled,noend]{algorithm2e}
\usepackage{lscape}
\usepackage{subfigure}
\usepackage{array}
\usepackage{pifont}
\usepackage{color, colortbl}
\usepackage{alltt}
\usepackage{listings}
\usepackage{balance}
\usepackage{multirow}

\DeclareFloatingEnvironment{plate}
\definecolor{Gray}{gray}{0.9}

\newcommand{\nurl}{nURL}
\newcommand{\nurls}{nURLs}
\newcommand{\toolname}{YourAdValue}

\definecolor{heraldBlue}{rgb}{0.0,0.0,0.8}
\definecolor{heraldRed}{rgb}{0.8,0.0,0.0}
\definecolor{heraldGray}{rgb}{0.4,0.4,0.4}
\definecolor{heraldBlack}{rgb}{0.0,0.0,0.0} 
\definecolor{heraldGreen}{rgb}{0.0,0.4,0.0} 

\newcommand{\ignore}[1]{}

\hypersetup{
  bookmarks=false, 
  colorlinks=true,%
  citecolor=black,%
  filecolor=black,%
  linkcolor=black,%
  urlcolor=black,%
  pagebackref=true,%
}

\def\ttfntsize{9}
\makeatletter
\let\oldtexttt\texttt
\let\texttt\@undefined
\makeatother
\newcommand{\texttt}[1]{\fontsize{\ttfntsize}{\ttfntsize}\oldtexttt{#1}}
\makeatletter
\let\oldtt\tt
\let\tt\@undefined
\makeatother
\newcommand{\tt}{\fontsize{\ttfntsize}{\ttfntsize}\oldtt}

\begin{document}

\copyrightyear{2017} 
\acmYear{2017} 
\setcopyright{acmcopyright}
\acmConference{IMC '17}{November 1--3, 2017}{London, United
Kingdom}\acmPrice{15.00}\acmDOI{10.1145/3131365.3131397}
\acmISBN{978-1-4503-5118-8/17/11}

\newcommand{\shortTitle}{How much do advertisers pay to reach you?}
\title{If you are not paying for it, you are the product:\\ {\normalfont \shortTitle}}

\author{Panagiotis Papadopoulos}
\affiliation{%
  \institution{FORTH-ICS, Greece}
}
\email{panpap@ics.forth.gr}

\author{Nicolas Kourtellis}
\affiliation{%
  \institution{Telefonica Research, Spain}
}
\email{nicolas.kourtellis@telefonica.com}

\author{Pablo Rodriguez Rodriguez}
\affiliation{%
  \institution{Telefonica Alpha, Spain}
}
\email{pablo.rodriguezrodriguez@telefonica.com}

\author{Nikolaos Laoutaris}
\affiliation{%
  \institution{Data Transparency Lab, Spain}
}
\email{nikos@datatransparencylab.org}


\begin{abstract}
Online advertising is progressively moving towards a programmatic model in which ads are matched to actual interests of individuals collected as they browse the web. Letting the huge debate around privacy aside, a very important question in this area, for which little is known, is: \emph{How much do advertisers pay to reach an individual?}

In this study, we develop a first of its kind methodology for computing exactly that -- the price paid for a web user by the ad ecosystem -- and we do that in real time. Our approach is based on tapping on the Real Time Bidding (RTB) protocol to collect cleartext and encrypted prices for winning bids paid by advertisers in order to place targeted ads. Our main technical contribution is a method for tallying winning bids even when they are encrypted. 
We achieve this by training a model using as ground truth prices obtained by running our own ``probe'' ad-campaigns. We design our methodology through a browser extension and a back-end server that provides it with fresh models for encrypted bids. 
We validate our methodology using a one year long trace of 1600 mobile users and demonstrate that it can estimate a user's advertising worth with more than 82\% accuracy.
\end{abstract}

\maketitle
\renewcommand{\shortauthors}{Panagiotis Papadopoulos et al.}
\title[\shortTitle]{If you are not paying for it, you are the product:\\ {\normalfont \shortTitle}}


\section{Introduction}\label{sec:introduction}
In today's data-driven economy, the amount of user data an IT company holds has a direct and non-trivial contribution to its overall market valuation~\cite{businessValuation}.
Digital advertising is the most important means of monetizing such user data.
It grew to \$194.6 billion in 2016~\cite{adspending}  of which \$108 billion were due to mobile advertising.
In fact, more and more companies rush to participate in this rapidly growing advertising business either as advertisers, ad-exchanges (ADXs), demand-side platforms (DSPs), data management platforms (DMPs), or all of the above.
For these companies to increase their market share, they need to deliver more effective and highly targeted advertisements.
A way to achieve this is through programmatic instantaneous auctions.
An important enabler for this kind of auctions is the Real-Time Bidding (RTB) protocol for transacting digital display ads in real time.
RTB has been growing with an annual rate of 128\%~\cite{RTBgrowth}, and currently accounts for 74\% of programmatically purchased advertising.
In US alone it created a revenue of \$8.7 billion in 2016~\cite{rtbRevenue1}.

Consequently, the collection of user personal data has become more aggressive and sometimes even intrusive~\cite{dataCollection,googlePixel}, raising a huge public debate around the tradeoffs between (i) innovation in advertising and marketing, and (ii) basic civil rights regarding privacy and personal data protection~\cite{publicUnrest1,publicUnrest2}.
These increasing privacy concerns, drew the attention of a significant body of research, which studied users' privacy loss in conjunction to existing user tracking techniques~\cite{Acar:2014:WNF:2660267.2660347,englehardtonline,Nikiforakis:2013:CME:2497621.2498133,Eckersley:2010:UYW:1881151.1881152,Leung:2016:YUA:2987443.2987456}, 
and proposed various defence mechanisms to users~\cite{Nikiforakis:2015:PDF:2736277.2741090,appsVsWeb,ksub}.
Still, there is an outstanding question that remains unaddressed by the related work in the area. 
This question concerns transparency and is the following:
\emph{Based on the exposed user personal data, how much do advertisers pay to reach an individual?}


Despite the importance of this question, it is surprising how little is known about it.
There exist several reports about the \emph{average} revenue per user (ARPU) from online advertising~\cite{twitterARPU, facebookARPU, googleARPU}, but ARPU, as its name suggests, is just an average.
It can be calculated by dividing the total revenue of a company by the number of its monthly active users.
Computing the revenue per \emph{individual} user is a completely different matter for which very limited work is available.

In particular, the FDVT~\cite{fdvt} browser extension can estimate the value of an individual user for Facebook, by tapping on the platform's ad-planner.
Another important prior work~\cite{lukasz2014selling-privacy-auction} leverages similarly the RTB protocol and specifically its final stage, where the winning bidder (advertiser) gets notified about the auction's charge price per delivered impression.
These charge prices were initially transmitted in cleartext and focused solely on them. 
However, more and more advertising companies use encryption to reduce the risk of tampering, falsification or monitoring from competitors.
This trend renders that method inapplicable for the current and future ad ecosystem, whose majority of companies will deploy charge price encryption.
In contrast to these works, our present method takes into account all the web activity of a user (not only on Facebook), and all RTB traffic, i.e., both cleartext and encrypted prices.


In this paper, our motivation is to enhance \emph{transparency} in digital advertising and shed light on pricing dynamics in its personal 
data-driven ecosystem. Therefore, we develop and evaluate a first of its kind methodology for enabling end-users to estimate in real time their actual cost for advertisers, even when the latter encrypt the prices they pay.
Designed as a browser extension, our method can tally winning bids for ads shown to a user and display the resulting amount as she moves from site to site in real time.

%
%
%
In summary, we make the following main contributions:
\begin{enumerate} [noitemsep,nolistsep,leftmargin=0.4cm]
\item We propose the first to our knowledge holistic methodology to calculate the overall cost of a user for the RTB ad ecosystem, using both encrypted and cleartext price notifications from RTB-based  auctions.
\item We study the feasibility and efficiency of our proposed method by analyzing a year-long weblog of 1600 real mobile users.
Additionally, we design and perform an affordable (a few hundred dollars cost) 2-phase real world ad-campaign targeting ad-exchanges delivering cleartext and encrypted prices in order to enhance the real-users' extracted prices.
We show that even with a handful of features extracted from the ad-campaign, our methodology achieves an accuracy $>82$\%.
The resulting ARPU is $\sim$55\% higher than that computed based on cleartext RTB prices alone.
Our findings challenge the related work's basic assumption~\cite{lukasz2014selling-privacy-auction} that encrypted and plain text prices are similar (we found encrypted prices to be $\sim$$1.7\times$ higher).
Finally, we validate our methodology by comparing our average estimated user cost with the reported per user revenue of major advertising companies.
\item Using lessons from the study, we design a system where the users, by using a Chrome browser extension, can estimate in real-time, in a privacy-preserving fashion on the client side, the overall cost advertisers pay for them based on their exposed personal information.
In addition, they can also contribute anonymously their impression charge prices to a centralized platform for further research.\\
\end{enumerate}

\noindent{\bf Paper Organization.} 
Section~\ref{sec:background} summarizes various key concepts of the RTB ecosystem and presents the main challenge and motivation of our work.
Section~\ref{sec:methodology} provides a high-level overview of our novel methodology and our price modeling engine.
Section~\ref{sec:dataset} presents an analysis of the dataset we use to bootstrap our modeling of encrypted prices.
Section~\ref{sec:price-modeling} presents in detail the effort to model RTB charge prices by executing probing ad-campaigns. These campaigns provide ground-truth data, which is used to train a machine learning classifier that can estimate encrypted prices in real-time at the browser of a user.
Section~\ref{sec:user-cost} puts all the pieces together and presents results on the overall monetary cost for displaying ads to users. 
Section~\ref{sec:related-work} covers related work while Section~\ref{sec:discussion} discuses various aspects of our work and concludes the paper.


\vspace{-0.1cm}
\section {Background on RTB}\label{sec:background}
RTB accounts for 74\% of programmatically purchased advertising, reaching a total revenue of \$8.7 billion in US~\cite{rtbRevenue1}, allowing advertisers to evaluate the collected data of a given user at real-time and bid for an ad-slot in the user's display. 
Next, we briefly cover the most important aspects of RTB auctions, key entities involved (\S~\ref{subsec:key-roles}), and how they are relevant to our study (\S~\ref{subsec:rtb-nurls}).

\begin{figure}[t]
\vspace{-0.2cm}
	\includegraphics[width=1.08\columnwidth]{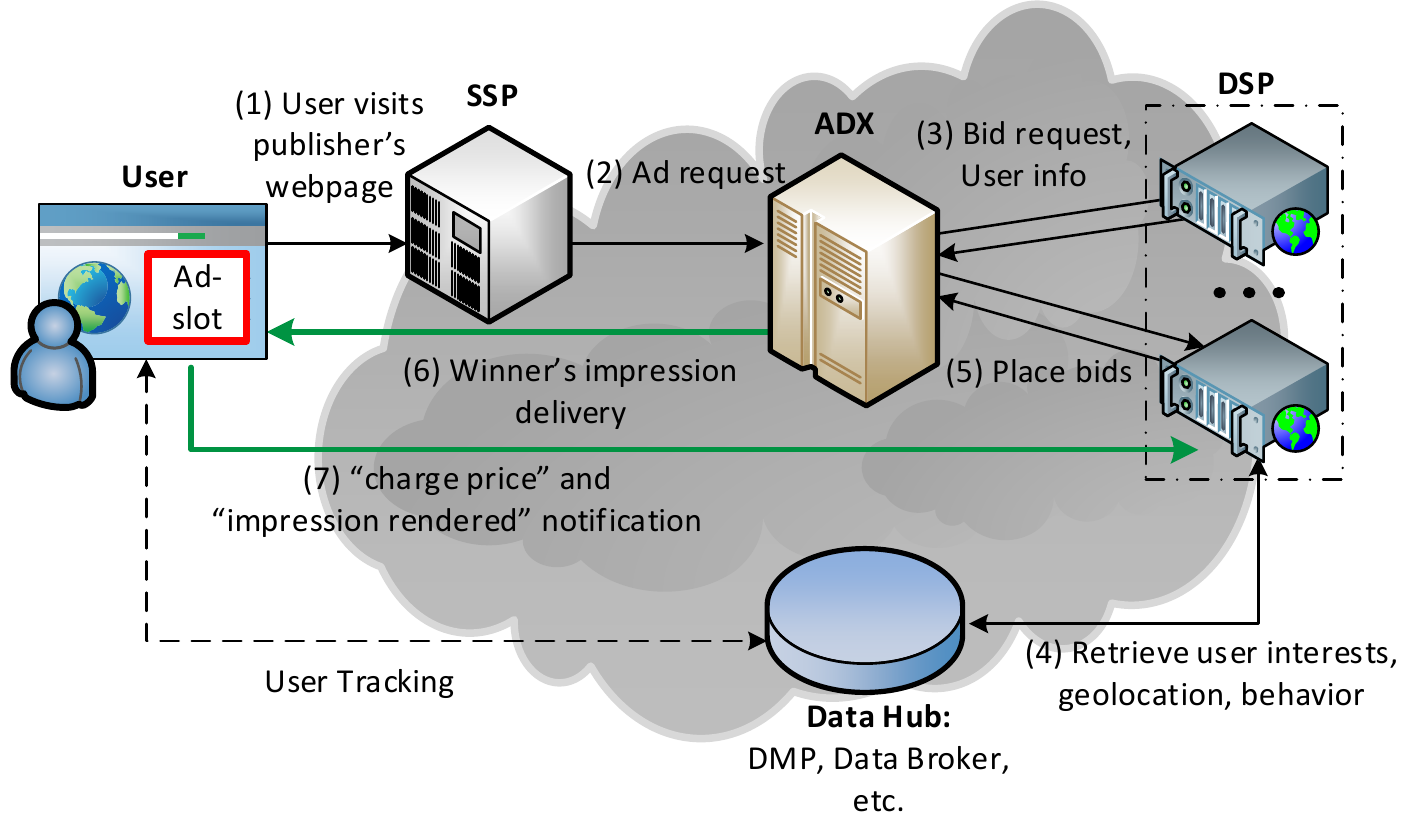} \vspace{-0.8cm}
	\caption{High level overview of the RTB ecosystem. Several entities interact with each other, exchanging user's personal data before it is finally converted to money.}
	\label{fig:rtb}
\end{figure}

\subsection{The key players}
\label{subsec:key-roles}

As it can be seen in Figure~\ref{fig:rtb}, the key roles of the RTB ecosystem include the \emph{Advertiser}, \emph{Publisher}, \emph{DSP}, \emph{Ad-exchange}, and \emph{SSP}, which interact with each other in several ways~\cite{bashirtracing}.
Note that it is very common for some (large) companies to play simultaneously different roles even inside the same auction (e.g. Google's DoubleClick Bid Manager~\cite{doubleclickbygoogle} and DoubleClick for Publishers~\cite{doubleclickformanagers}.

\noindent
{\bf Publisher:} (e.g., CNN.com) is the owner of a website, where users browse for content and receive ads (step 1).
Each time a user visits the website, an auction takes place for each available ad slot.
The ad impression of the winning advertiser is finally displayed in each auctioned slot of the website.

\noindent
{\bf Advertiser:} is the buyer of a website's ad slots. The advertiser creates ad campaigns and defines the audience that has to be targeted according to his marketing objectives, budgets, strategies, etc.
In each auction, the one with the highest bid wins the ad slot and places its impression on the screen of the website's visitor.

\noindent
{\bf Supply-Side Platform (SSP):} is an agency platform, which enables publishers to manage their inventory of available ad slots and their 
pricing, allocate ad impressions from different channels (e.g. RTB or backfill in case of unsold inventory~\cite{backfill}) and receive revenue\footnote{Publishers can also interface directly with ADXs and handle their inventory on their own.}.
SSP is also responsible for interfacing the publisher's side to multiple ad-exchanges (step 2) and aggregate/manage publisher's connections with multiple ad networks and buyers.
In addition, by using web beacons and cookie synchronization, SSPs perform user tracking in order to better configure their ad slots' pricing and achieve as many re-targeting ads as possible and thus higher bids~\cite{sspUserData}.
Popular vendors selling SSP technology are OpenX, PubMatic, Rubicon Project, Right Media.

\noindent
{\bf Ad-exchange (ADX):} is a digital, real-time marketplace that, similarly to a stock exchange, enables advertisers and publishers to buy and sell advertising space through RTB-based auctions.
ADX is responsible for hosting an RTB-based auction and distribute the ad requests along with user information it owns (i.e. browsing history, demographics, 
location, cookie-related info) among all the interested auction participants (step 3).

Typically these auctions follow the second higher price model (i.e. Vickrey auctions)~\cite{vickrey1961counterspeculation}, thus, the charge price for the winner of the slot is the second highest submitted.
After the auction, the winning impression is served to the user's display within 100 ms of the initiating call (step 6) and the winning bidder is notified about the final charge price.
Popular ad-exchanges include: DoubleClick, MoPub, and OpenX.

\begin{table}[t]
	{\small
		\begin{tabular}{l}
			{\bf Winning Price Notification URLs} \\
			\toprule
			{\bf (A)} cpp.imp.mpx.mopub.com/imp?ad\_domain=amazon.es\& \\
			ads\_creative\_id=ID\&\textbf{\textcolor{heraldBlue}{bid\_price=0.99}}\&bidder\_id=ID\&... \\
			\&bidder\_name=..\&\textbf{\textcolor{heraldBlue}{charge\_price=0.95}}\&country=..\&... \\
			\&currency=USD\&latency=0.116\&mopub\_id=ID\&pub\_name=.. \vspace{0.3cm} \\ 
			
			{\bf (B)} tags.mathtag.com/notify/js?exch=ruc\&... \\
			\&\textbf{\textcolor{heraldBlue}{price=B6A3F3C19F50C7FD}}\&... \\
			\&3pck=http\%3A\%2F\%2Fbeacon-eu2.rubiconproject.com\%2F \\
			beacon\%2Ft\%2Fce48666c-6eb4-46db-b0e9-6f4155eb557d\%2F\vspace{0.3cm} \\ 
			
			{\bf (C)} adserver-ir-p.mythings.com/ads/admainrtb.aspx?googid=ID\&.. \\
			\&width=300\&height=250\&...\&cmpid=ID\&gid=ID\&mcpm=60\&... \\
			\textbf{\textcolor{heraldBlue}{rtbwinprice=VLwbi4K21KFAAAm2ziqnOS\_O5oNkFuuJw}}\&.. \\
			\bottomrule
		\end{tabular}
	}
	\caption{Examples of (A) cleartext, (B, C) encrypted RTB price notifications. ``ID'' is typically a hexadecimal number. }
	\label{tbl:nurls}\vspace{-0.6cm}
\end{table}

\noindent
{\bf Demand-Side Platform (DSP):} is an agency platform, which employs decision engines with sophisticated audience targeting and optimization algorithms aiming to help advertisers buy the best-matched ad slots from ADXs in a simple, convenient and unified way.
DSPs retrieve and process user data from several sources (step 4) such as ADXs, Data Hubs, etc.
The result of this processing is translated to a decision in practice: \emph{How much is it worth to bid for an ad slot for this user, if any?}
If the visitor's profile matches the audience the advertiser has focused his ad-campaign on, the DSP will submit to the ADX the impression and a bid in CPM (cost per 1000 impressions~\cite{cpm}, typically in USD or Euro) on behalf of the advertiser (step 5).
Popular DSPs are MediaMath, Criteo, DoubleClick, AppNexus and Invite Media.

\noindent
{\bf User Data Hub, Data Exchange Platform (DXP):} is a centralized data warehouse such as a Data Management Platform (DMP)~\cite{dmp,dmp2} or a Data Broker~\cite{dataBroksers} which aggregates, cleans, analyzes and maintains user private data such as demographics, device fingerprints, interests, online and offline contextual and behavioral information~\cite{offlineUserData,offlineUserData2}.
These user data are typically aggregated in two formats: 1) a full, audience user profile for offline analytics and data mining, 2) a run-time user profile, optimized for real-time requests such as RTB queries from DSPs, before submitting their bids to ADXs~\cite{turn,dmpAnalytical} (step 4).

Such user profiles are sold to ad entities~\cite{dxp} because they increase the value of an RTB inventory by enabling a more behavioral-targeted advertising ($2.7\times$ more effective than non-targeted advertising~\cite{beales2010value,Yan:2009:MBT:1526709.1526745}).
In fact, Data Hubs are considered the core component of the digital ad-ecosystem as they perform the attribution and labeling of users' data and create groups, namely \emph{audience segments}, which are useful (i) to the publishers for their customer understanding, (ii) to the SSPs for retrieving more re-targeted ads and (iii) to the DSPs for feeding their bid decision engines.
Further, quality scores are impartially assigned to users' private data based on the success of ad-campaigns they were used, thus driving the bid prices of future ad-campaigns.
Notable DXPs are Turn, Adobe, Krux, Bluekai, Lotame.

%

\subsection{RTB price notification channel}
\label{subsec:rtb-nurls}

When an ADX selects the winning bid of an auction, the corresponding bidder must be notified about its win to log the successful entry and the price to be paid to the ADX.
One could implement this notification in two ways: (i) with a server-to-server message between ADX and DSP, (ii) with a notification message conjoined with the price, passed through the user's browser as a call-back to the DSP.

The first option is straightforward and tamper-proof; no one can modify or block these messages, allowing companies to ensure that their logs are fully synced at any time.
In addition, DSPs can hide information about the transactions, the purchased ad-slots and the prices paid from the prying eyes of competitors.
However, DSPs do not have any indication of the delivery of each ad, in order to inform their campaigns and budget. 

Instead, the second option not only can ensure the DSP that the winning impression was indeed delivered (the callback is fired soon after the impression is rendered on the user's device), but also gives the opportunity to drop a cookie on the user's device.
Therefore, the second option is the dominant one in the current market: the ADX piggybacks a notification URL (\nurl) in the ad-response, which delivers to the user the winning impression and the ad (steps 6 and 7 in Figure~\ref{fig:rtb}).
This \nurl\ includes basically the winning DSP's domain, the charge price, the impression ID, the auction ID and other relevant logistics (see Table~\ref{tbl:nurls} for some examples).
In this present work, we study such \nurls\ and the prices embedded in them, as well as how they associate with the users' browsing behavior and other personal information.




\begin{figure}[t]
	\begin{minipage}{0.242\textwidth}
	\centering
	\vspace{-0.2cm}
	\includegraphics[width=1.09\columnwidth]{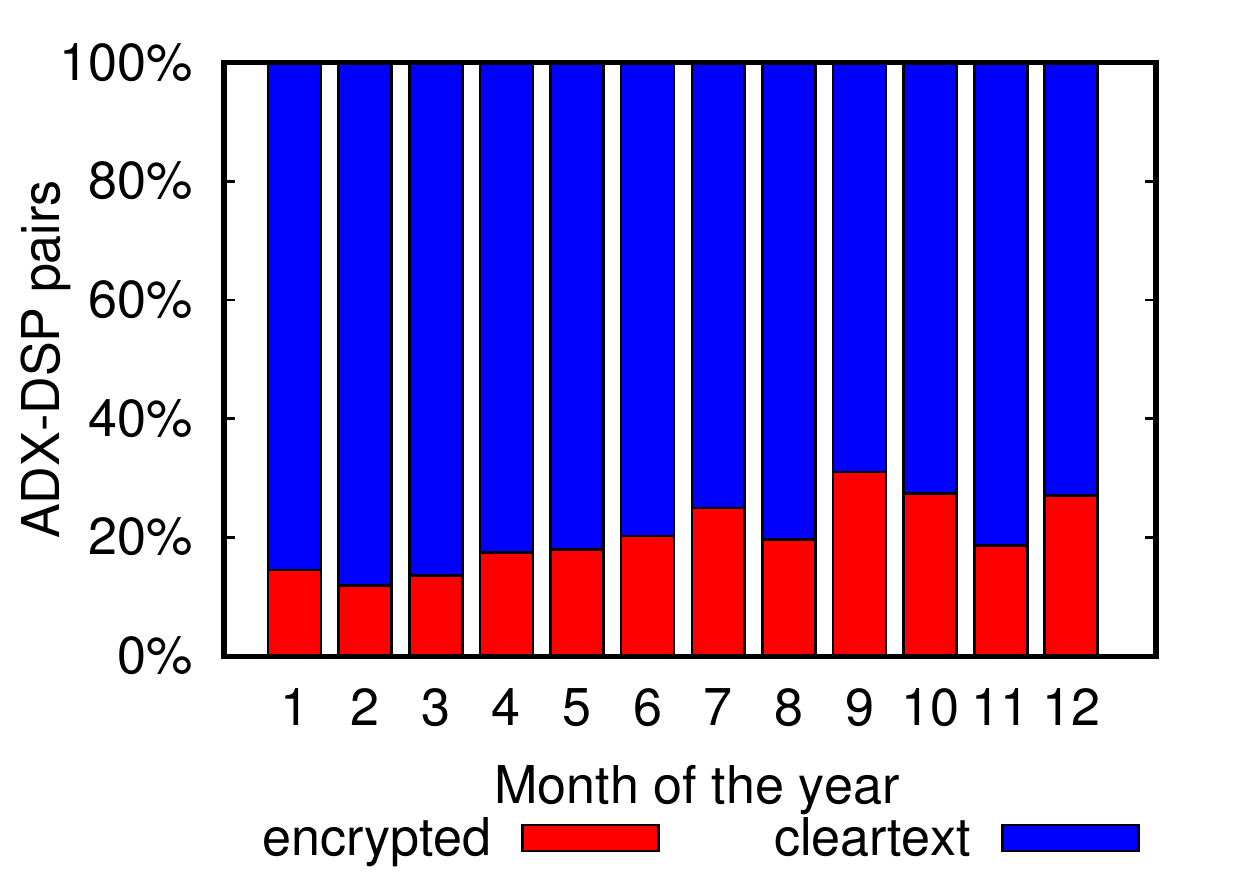}	\vspace{-0.5cm}
	\caption{Portion of encrypted and cleartext pairs of ADX-DSP over time (2015).}
	\label{fig:encryptedpairs}
	\end{minipage}
\hfill
	\begin{minipage}{0.227\textwidth}
	\centering
	\includegraphics[width=1.09\columnwidth]{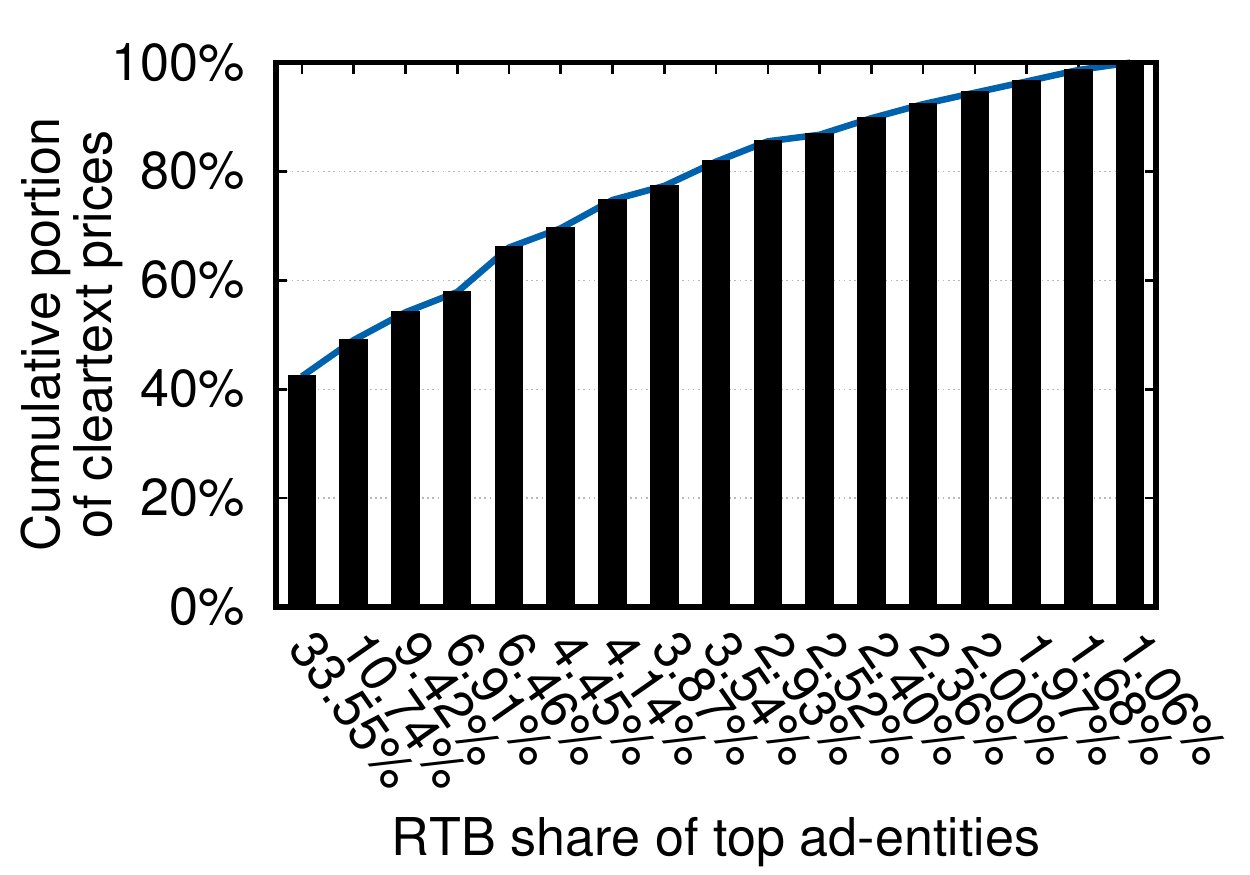} \vspace{-0.5cm}
	\caption{Cumulative portion of cleartext prices vs. ad-entities' portion of RTB.}
	\label{fig:flipToEncrypted}
		\end{minipage}\vspace{-0.1cm}
\end{figure}

\subsection{Encrypted vs. cleartext prices}
Although in the early years of RTB, all charge prices in \nurls\ were in cleartext, we see that nowadays more and more
companies deliver charge prices in encrypted form (see examples in Table~\ref{tbl:nurls}). 
While cleartext prices captured at the user's browser can be easily tallied to estimate the total cleartext cost, the same does not apply for the encrypted prices.
The popular 28-byte encryption scheme companies use cannot be easily broken~\cite{decryptPrices}. 

Previous studies~\cite{lukasz2014selling-privacy-auction} assumed that encrypted prices follow the same distribution as cleartext ones.
Indeed, one may argue that the price encryption is just to avoid tampering of reported prices, so encrypted prices probably follow the cleartext price distribution.
However, encryption provides also confidentiality to the bidding strategy.
Thus, possible use of encryption in charge prices may be also a sign of a higher value that the bidder wants to hide: aggressive re-targeting because of user's previous incomplete purchases, targeting users with higher spending habits, or users with specialized needs (e.g., sensitive products, expensive drugs, etc.).
Hence, a bidder (e.g. a DSP) may choose encryption to reduce transparency over its bidding strategies, or possible special knowledge it may have about a specific user, thus preventing an external observer or competitor from assessing its bidding methods and ad-campaigns.

We should note that encryption is not a feature that comes for free.
There are significant costs for the participating parties such as more computation and storage overhead, energy consumption and higher imposed latency.
Therefore, these costs alone could be a reason for an ADX to charge more for providing the benefits of encryption to a DSP.
Considering all the aforementioned, in our study, we remove the need for making any assumptions regarding encrypted prices and allows us to account for any potential differences between cleartext and encrypted prices.

\begin{figure}[t]
	\includegraphics[width=1.04\columnwidth]{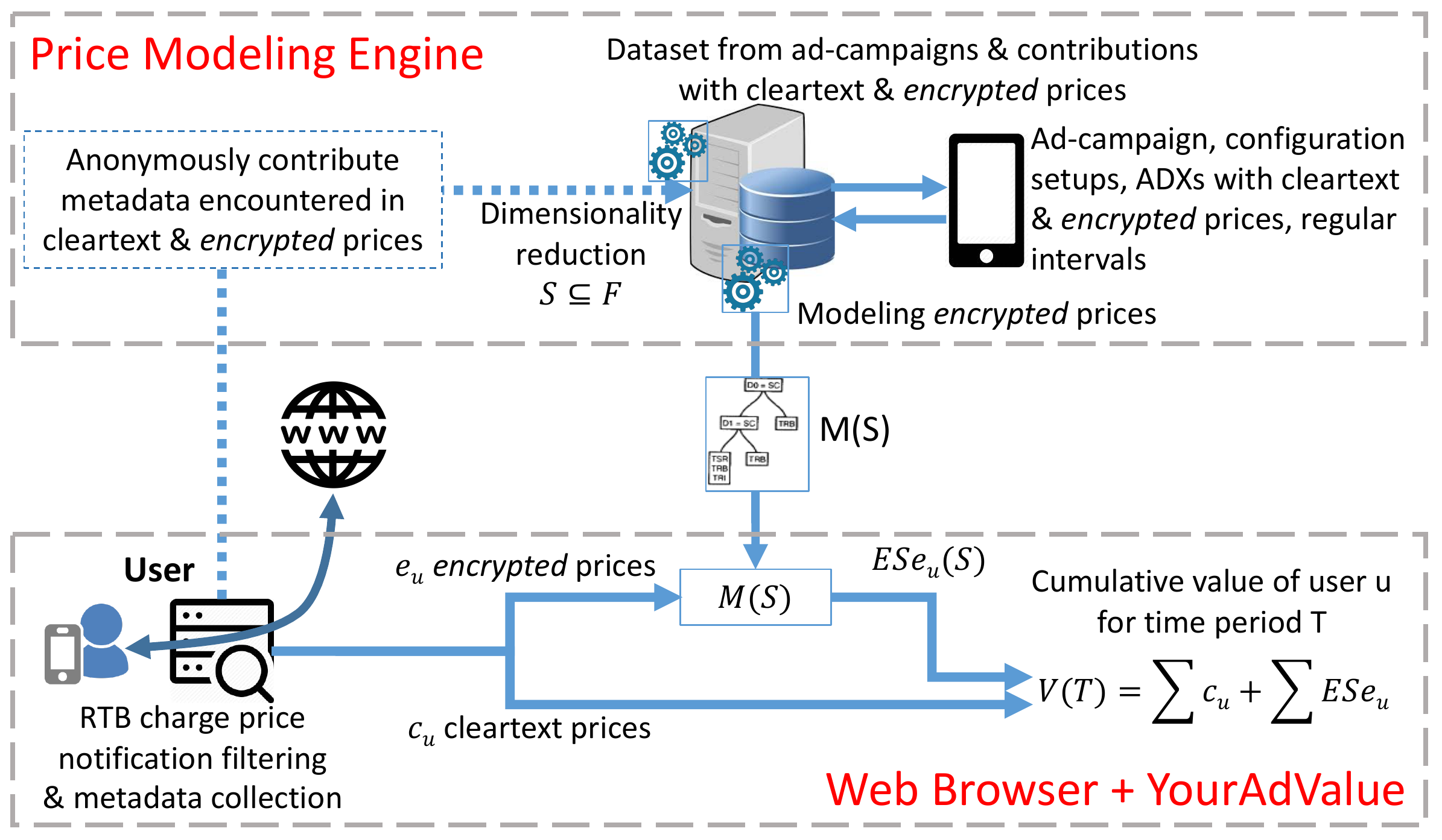} \vspace{-0.7cm}
	\caption{High level overview of our method. The user deploys \toolname\ on her device, which calculates in real-time the total cost paid 
	for her by advertisers. In case of encrypted prices, it applies a decision tree model derived from the PME.}
	\label{fig:system}
\end{figure}

\subsection{Encrypted prices on the rise}
Encryption is a regular practice in desktop RTB ads ($\sim$$68\%$ as reported in~\cite{olejnikbid} with major supporters being DoubleClick, RubiconProject and OpenX).
By analyzing a weblog of 1600 real mobile users (see Section~\ref{sec:dataset}), we detected a smaller portion in mobile RTB ads ($\sim$$26\%$).
However, we found that the percentage of ADX-DSP pairs using encrypted price \nurls\ was steadily increasing through time (Figure~\ref{fig:encryptedpairs}), which means that more and more mobile advertising entities have started using \nurls\ with encrypted prices.

In fact, we found that the mobile advertising entities with the larger RTB shares deliver the highest portion of cleartext prices as well (Figure~\ref{fig:flipToEncrypted}).
For example, MoPub and Adnxs, the two leading ad-entities in our dataset, are responsible for $33.55\%$ and $10.74\%$ of the overall RTB ads detected, respectively (x-axis).
They are also responsible for $45.40\%$ and $5.45\%$ for the cleartext prices detected, respectively (cumulatively in y-axis).
If these two (or more) companies flipped their strategy from cleartext to encrypted, it would dramatically impact the RTB-ecosystem's transparency and hinder price information exposed to an external auditor or the involved user.

Given these trends in mobile and desktop, we expect that in the near future RTB auctions will dominate, and many of the ad-entities will use encryption to deliver their charge prices.
Our methodology anticipates these trends and promotes better transparency in online advertising and usage of user personal data.
This methodology allows end-users to accurately estimate on their browser, at real-time, their average ad-related cumulative cost, even when the charge prices are encrypted.

\section{Methodology}\label{sec:methodology}
In this section, we describe our proposed methodology, with which a user $u$ can estimate in real-time the accumulated cost $V_u$ for the ads she was delivered while browsing the web (\S~\ref{subsec:total-cost}) (notations used are summarized in Table~\ref{tbl:notations}).
Following this methodology, we design our system based on two main components: (i) a remote \emph{Price Modeling Engine} (\S~\ref{subsec:pme}) and (ii) a user-side tool, namely \emph{\toolname} (\S~\ref{subsec:plugin}).
Figure~\ref{fig:system} presents an overview of our proposed methodology.

\begin{table}[t]
	\centering
	\begin{small}
		\begin{tabular}{p{1.5cm}|l}
			{\bf Notation}		&	{\bf Definition}								\\
			\hline
			$V_u$			&	Total cost of user $u$						\\
			$C_u$, $E_u$		&	Sum of cleartext, encrypted prices of user $u$		\\
			$SC_u$, $SE_u$	&	Set of cleartext, encrypted price \nurls\ of user $u$	\\
			$F_i$			&	Vector of features for a price \nurl\ $i$			\\
			$S_i \subseteq F_i$	&	Core features selected to represent \nurl\ $i$		\\
			$ESe(S_i)$		&	Estimated encrypted price based on vector		\\
							&	of features $S$ of price \nurl\ $i$				\\
		\end{tabular}
	\end{small}
	\caption{Summary of notations.}
	\label{tbl:notations}
\end{table}

\subsection{Overall cost of the user's data}
\label{subsec:total-cost}

The overall ad-cost of the user for time period $T$ is the sum of charge prices the advertisers have paid after evaluating her personal data they have collected and delivering ads to her device.
Specifically, this overall value is the sum of both her cleartext $C_u(T)$ and encrypted $E_u(T)$ prices and can be stated as:
~
\begin{equation}
	V_u(T)=C_u(T) + E_u(T)
	\label{form:total}
\end{equation}

\noindent
The cleartext prices of a user can be aggregated in a straightforward fashion, thus producing the ad-cost for user $u$ over such prices:
\begin{equation}
	C_u(T)=\sum\limits_{i} c_i,\hspace{0.1cm}where \hspace{0.2cm} i \in SC_u(T)
	\label{form:cleartext}
\end{equation}

On the other hand, the calculation of the aggregated $E_u(T)$ of the encrypted prices for the same user is not easy.
The actual price values $e_i$ are hidden and therefore need to be estimated.
To achieve that, we leverage the metadata of each charge price in the user's set $SE_u$(T) of encrypted price notifications.
Such metadata may include: time of day, day of week, size of ad, DSP/ADX involved, location, type of device, associated IAB, type of OS, user's interests, etc.
All these metadata are collected in a feature vector $F_i$ that captures the context of a specific charge price $e_i$ in \nurl$i$.

In order to estimate each encrypted notification price $i$, we built a machine learning model, which receives as input the feature vector $F_i$ (or a subset $S_i \subseteq F_i$), extracted from the \nurl$i$, and estimates a charge price $ESe(S_i)$ for the encrypted price $e_i$.
This permits us then to aggregate the estimated encrypted prices for user $u$ as we have done for the cleartext ones:
~
\begin{equation}
E_u(T)=\sum\limits_{i} ESe(S_i),\hspace{0.1cm}where \hspace{0.2cm} i \in SE_u(T)
\label{form:encrtypted}
\end{equation}

\subsection{Price Modeling Engine}
\label{subsec:pme}

The core element of our solution, the Price Modeling Engine (PME), is a centralized repository responsible for the estimation of encrypted prices. To achieve this, the PME requires a sample of charge price data and associated features to train a machine learning model.
This component is designed to incorporate data such as offline weblogs (see Section~\ref{sec:dataset}), or online anonymous contributions (anonymized features and charge prices) from participating users, similarly to other systems that depend on crowd-sourcing (e.g., Floodwatch~\cite{floodwatch}).
Using such data, the PME can re-train the computed model at any time.
To assess the difference between cleartext and encrypted price distributions in the wild and fine-tune the training model, the PME runs small ``probing ad-campaigns'' to collect ground truth of real charge prices from both encrypted and cleartext formats.

Feeding the PME with all possible metadata available, i.e. auctions' metadata and users' personal data, is clearly not practical. There exist hundreds of data points per individual price. Passing all of them to the modeling engine would make the computational cost excessive.
Additionally, if all data points were to be used in the probing ad-campaigns, they would render such campaigns too expensive for their purpose.
In order to run effective and efficient ad-campaigns, and allow the training of a price model without high computation overhead, the PME performs careful dimensionality reduction on the extracted metadata ($F$) to derive a subset $S\subseteq F$ of core features capable to capture the value of an impression.
This dimensionality reduction makes the probing ad-campaigns feasible by reducing by many orders of magnitude the needed features of each testing setup, and effectively the number of setups to be tested (see Section~\ref{sec:price-modeling}).

Using the collected ground truth of encrypted prices from ad-campaigns, the PME trains a machine learning model $M$ to infer encrypted prices based on their associated subset of features $S$.
Then, each user can apply the model $M$ (in the form of a decision tree) locally on their device to estimate each of her encrypted charge prices based on the matching metadata $S$.

In case the availability of cleartext prices is limited, the reduction step to identify important features could be hindered, but not obstructed.
To mitigate this, the PME can run more probing ad-campaigns to cover extra features found in users' anonymous contributions, or that are available in professional ad-campaign planners (as in FDVT~\cite{fdvt}).
Then, the most important features can be selected based on their contribution to model the encrypted prices extracted from these campaigns.

\begin{table}[t]
	\centering
	\begin{small}
		\begin{tabular}{l|c|c|c}
			{\bf Metric}		&	{\bf D}	&	{\bf A1 }		&	{\bf A2 }	\\
			\hline
			Time period & 12 months& 13 days	& 8 days	\\  \hline
			Impressions	& 78560	&	632667	& 318964 \\ \hline
			RTB	publishers & $\sim$5.6k/month	& $\sim$0.2k & $\sim$0.3k	\\ \hline
			IAB categories & 18 & 16 &	7 \\ \hline
			Users &	1594 	& -	& - \\ \hline
		\end{tabular}
	\end{small}
	\caption{Summary of dataset and ad-campaigns.} 
	\label{tbl:summaries}
\end{table}
\subsection{ \toolname}
\label{subsec:plugin}

\toolname\ is a user-side tool responsible for monitoring the user's \nurls\ and calculating locally the cumulative cost paid for her in real-time.
To achieve this, it filters \nurls\ from her network traffic and extracts (i) the RTB auction's charge prices (both encrypted and cleartext), and (ii) metadata from each specific auction (e.g. time of day, day of week, size of ad, involved DSP and ADX, etc.) along with the personal data the user leaks while using online services (e.g. location, type of device and browser, type of OS, browsing history, etc.).

As we mentioned earlier, cleartext prices can be aggregated directly, but encrypted prices must be estimated.
Therefore, \toolname\ retrieves from the PME a model $M(S_i)$ that (i) includes the features $S_i$ that need to be extracted from the collected metadata, and (ii) provides a decision tree for the estimation of an encrypted price based on these features.

Using this model, \toolname\ can estimate locally on the client side, the value $ESe(S_i)$ of the encrypted charge prices based on the features $S_i$ of the given \nurl.
After estimating each encrypted price, \toolname\ presents to the user the calculated sums $C_u(T)$ and 
$E_u(T)$ along with relevant statistics and the total amount $V_u(T)$ paid by advertisers (see Section~\ref{sec:user-cost}).

\toolname\ can be implemented in the same manner, either as a browser extension for desktops or as a module for mobile devices.
In the latter case, \toolname\ can monitor traffic of both browsers and apps similar to existing approaches~\cite{appsVsWeb}.
For simplicity, in this work we design \toolname\ as a browser extension; its mobile counterpart is part of our future work.

Our tool, built as an extension for Chrome browser, monitors both HTTP and HTTPS traffic of the user and detects the RTB \nurls.
Additionally, it stores in the browser's local storage the filtered charge prices, the personal and auctions metadata and the estimation of the encrypted prices.
The extension, through toolbar notifications, informs the user about newly detected RTB charge prices. 
Upon request, it reports the cumulative cost along with the previous individual charge prices.
Finally, the extension periodically issues requests to PME to check for new versions of the model.

\begin{table}[t]
	\begin{small}
		\begin{tabular}{p{1.7cm}|p{6.2cm}}
			{\bf Type} & {\bf Feature}	\\
			\hline
			\multirow{3}{*}{\bf Geo-temporal}	&	Time of day, Day of week												\\\cline{2-2}
										&	Location of user based on IP, \# of unique locations of the user, location history \\\hline
			\multirow{9}{*}{\bf User}			&	Interest categories of the user, Type of mobile device, \# of total web beacons detected for the user, \# of cookie syncs detected of the user up to now, \# of publishers visited by the user, \# of total bytes consumed by the user,					\\\cline{2-2}
										&	Avg. number of reqs per user for the advertiser, \# of HTTP reqs of the user, Avg. number of bytes per req of user, Total duration of reqs of the user, Avg. duration per req of the user														\\\hline
			\multirow{7}{*}{\bf Ad}			&	Size of ad, ADX of \nurl, DSP of \nurl, IAB category of the publisher, popularity of particular ad-campaign,\\\cline{2-2}
										&	\# of total HTTP reqs of the advertiser, \# of bytes of HTTP req, Avg. duration of the reqs for the advertiser, \# of URL parameters, Number of total bytes delivered for the advertiser\\  \hline
		\end{tabular}
		\caption{Features extracted by summarizing data from parameters embedded in each price notification detected in the dataset for users and advertisers.}
		\label{tab:aggregated-features}
	\end{small}
\end{table}


\begin{figure*}[t]
	\centering
	\begin{minipage}{0.32\textwidth}
		\centering\vspace{-0.1cm}
		\includegraphics[width=1\columnwidth]{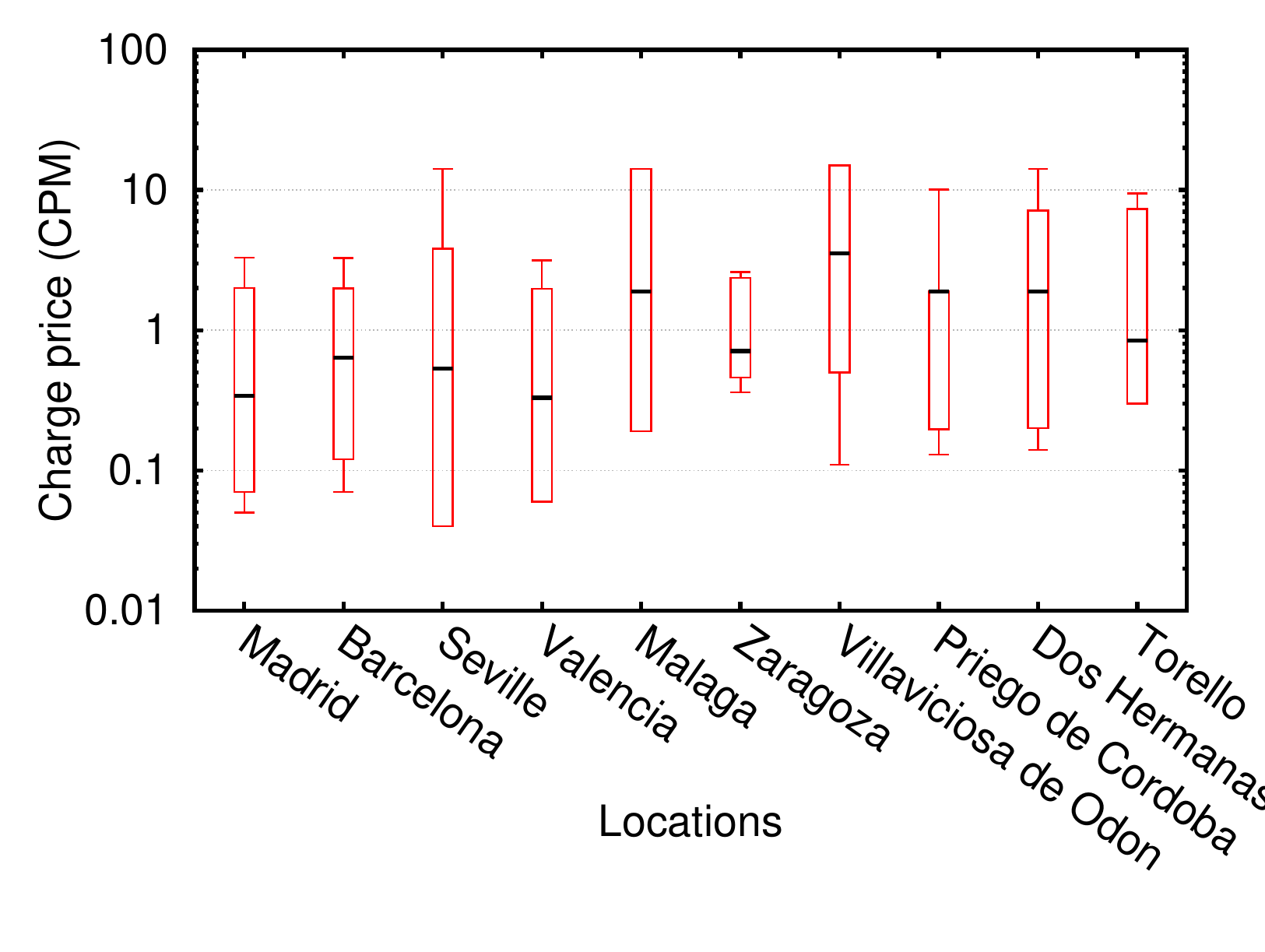}
				\vspace{-1cm}
		\caption{Distribution of charge prices per city (sorted by city size).}
		\label{fig:locationsWiskers}
	\end{minipage}
	\hfill
	\begin{minipage}{0.32\textwidth}
		\centering
		\vspace{-0.1cm}
		\includegraphics[width=1\columnwidth]{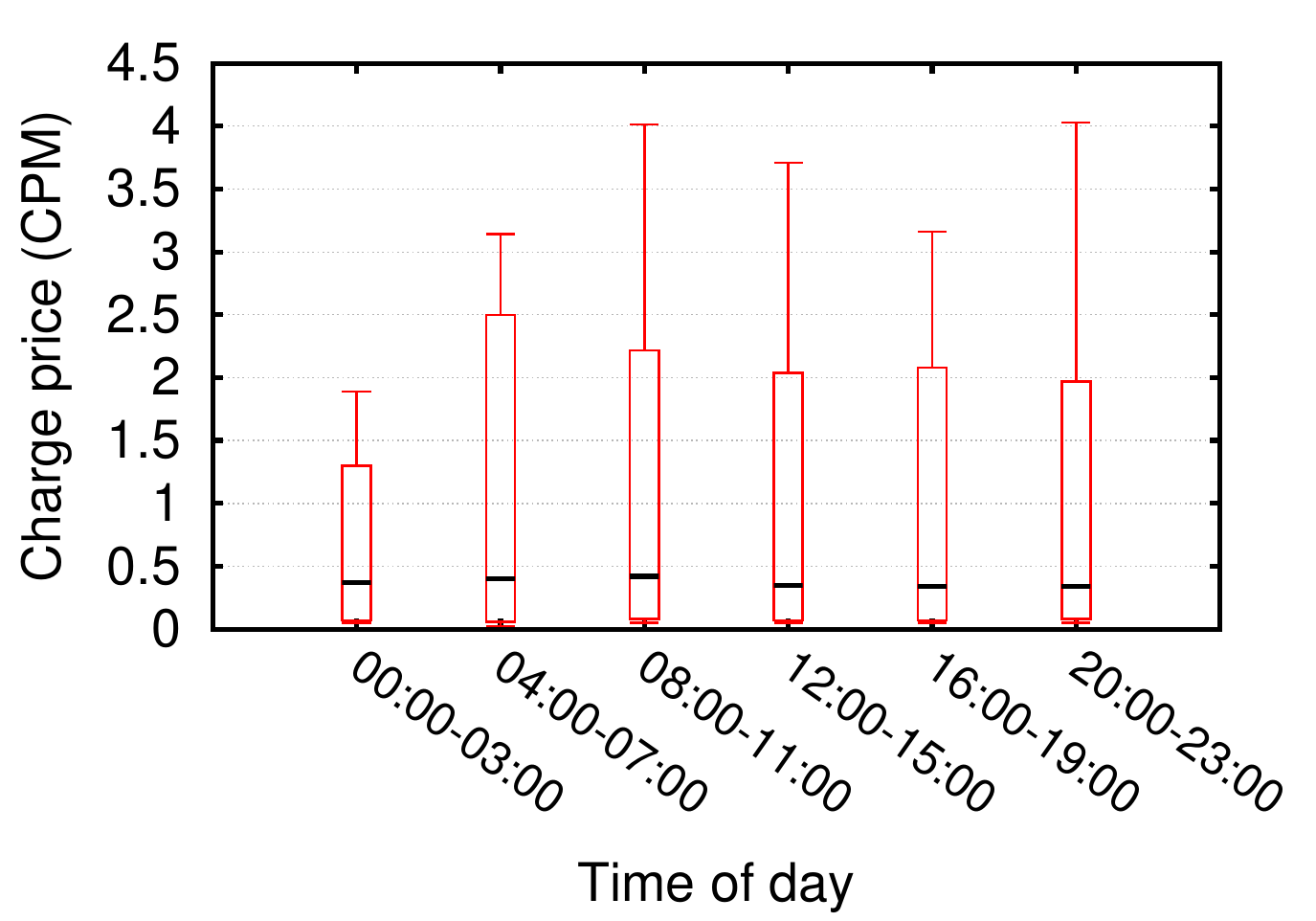}
				\vspace{-0.7cm}
		\caption{Distribution of charge prices for different times of day.}
		\label{fig:TODWiskers}
	\end{minipage}
	\hfill
	\begin{minipage}{0.32\textwidth}
		\centering
		\vspace{-0.1cm}
		\includegraphics[width=1\columnwidth]{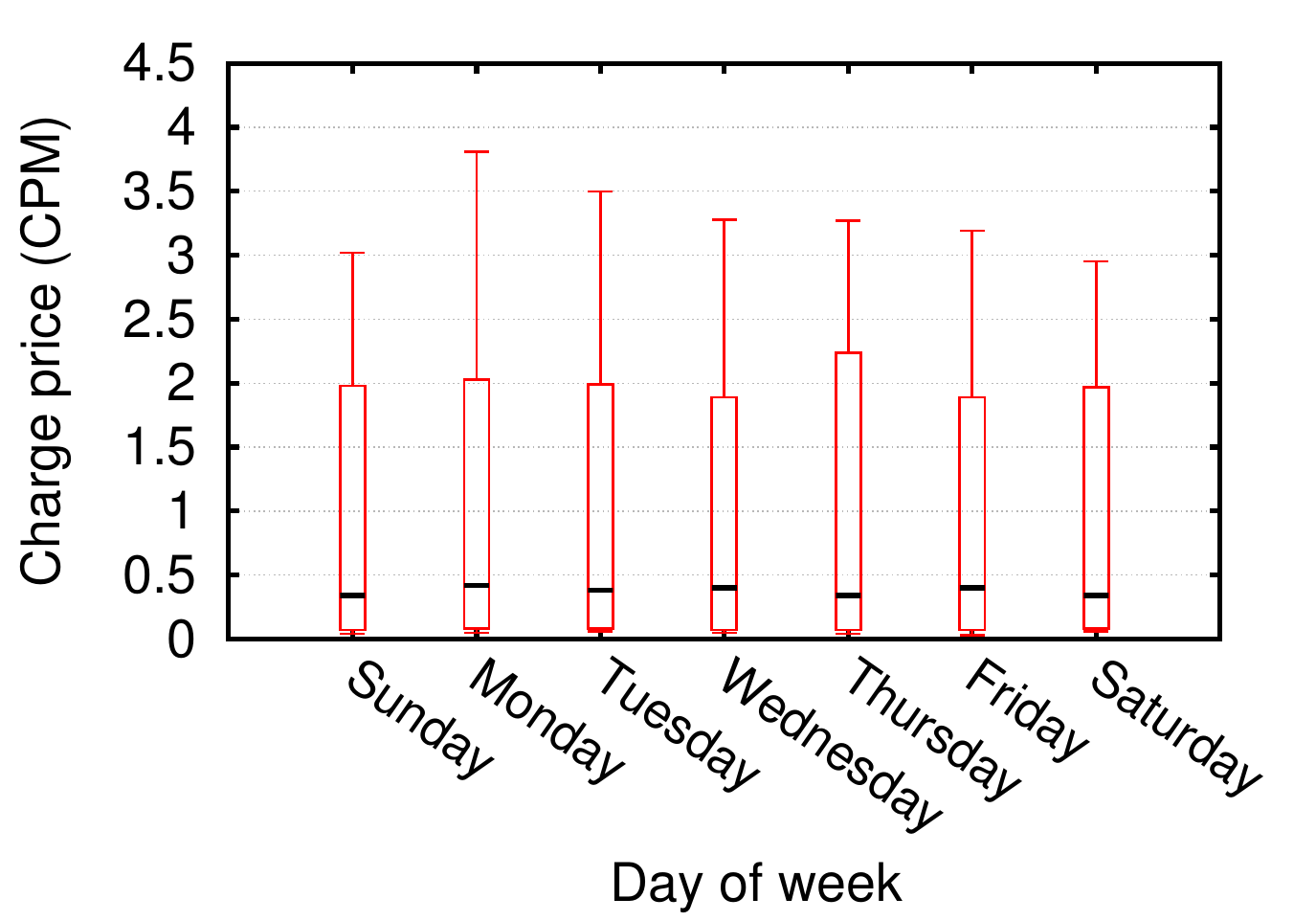}
				\vspace{-0.7cm}
		\caption{Distribution of charge prices for different days of week.}
		\label{fig:DOWWiskers}
	\end{minipage}
\end{figure*}
\begin{figure*}[t]
	\begin{minipage}{0.235\textwidth}
		\centering
		\includegraphics[width=1.1\columnwidth]{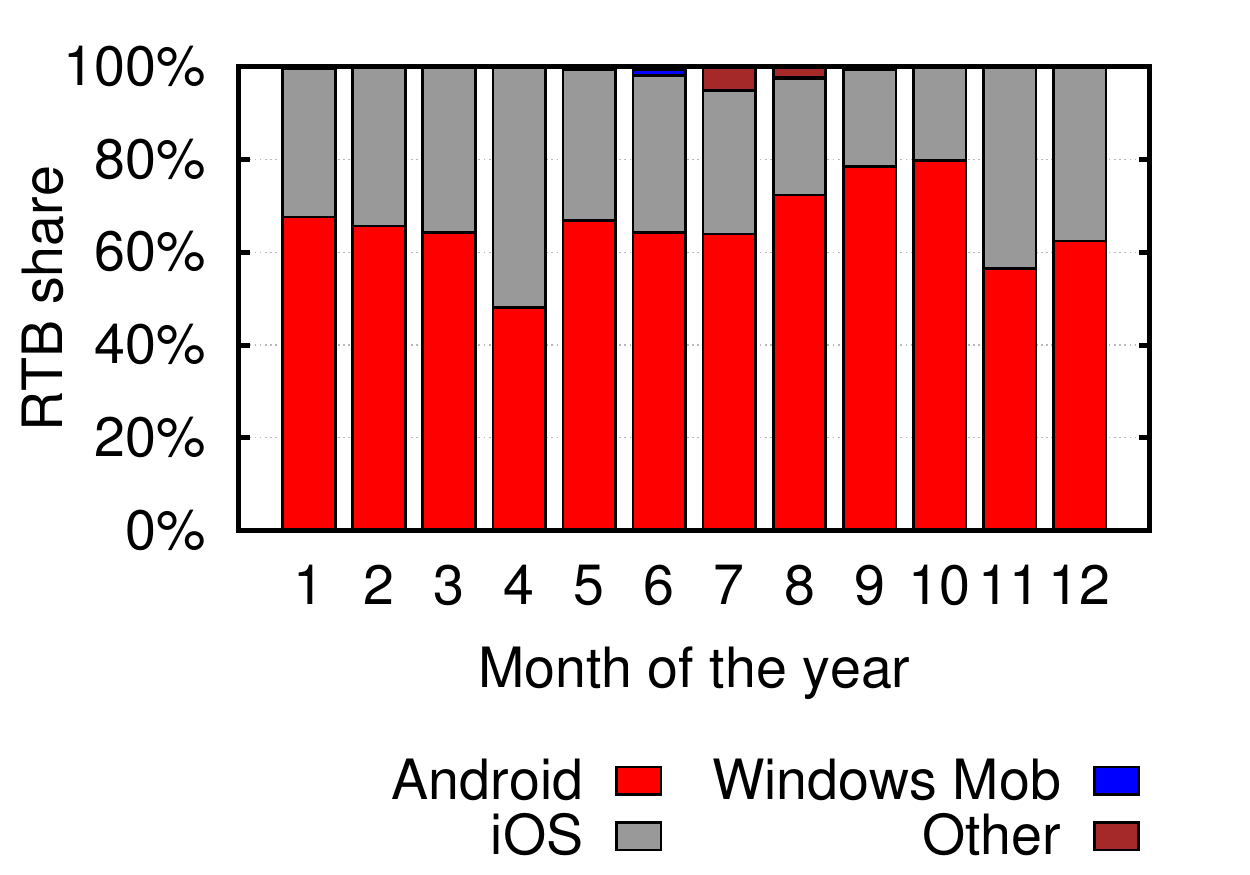}
		\caption{Portion of RTB traffic for top mobile OSes.}
		\label{fig:devicePopularity}
	\end{minipage}	
	\hfill	
	\begin{minipage}{0.235\textwidth}
		\centering
		\includegraphics[width=1.1\columnwidth]{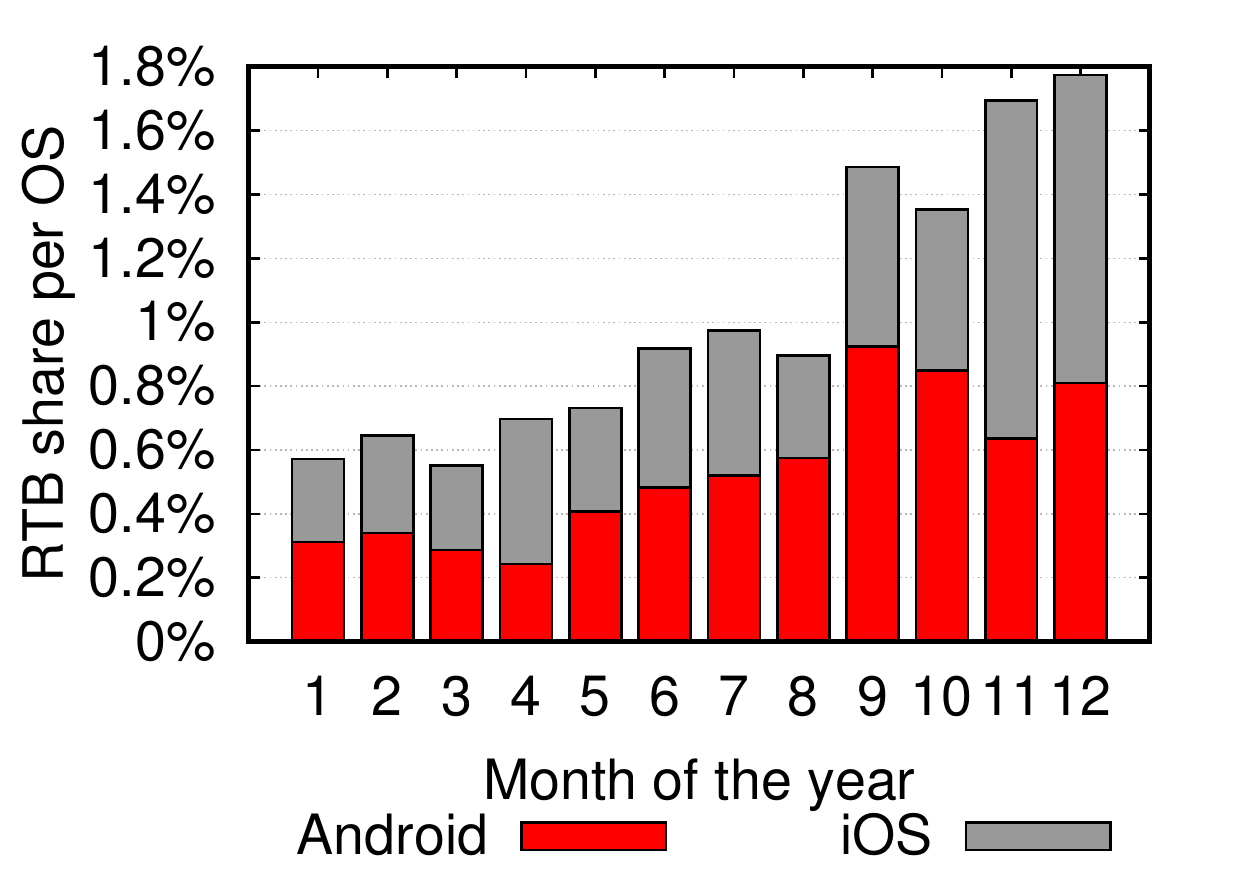}
		\caption{Portion of RTB traffic normalized by OS.}
		\label{fig:normPopularity}
	\end{minipage}
	\hfill	
	\begin{minipage}{0.23\textwidth}
		\centering
		\includegraphics[width=1.1\columnwidth]{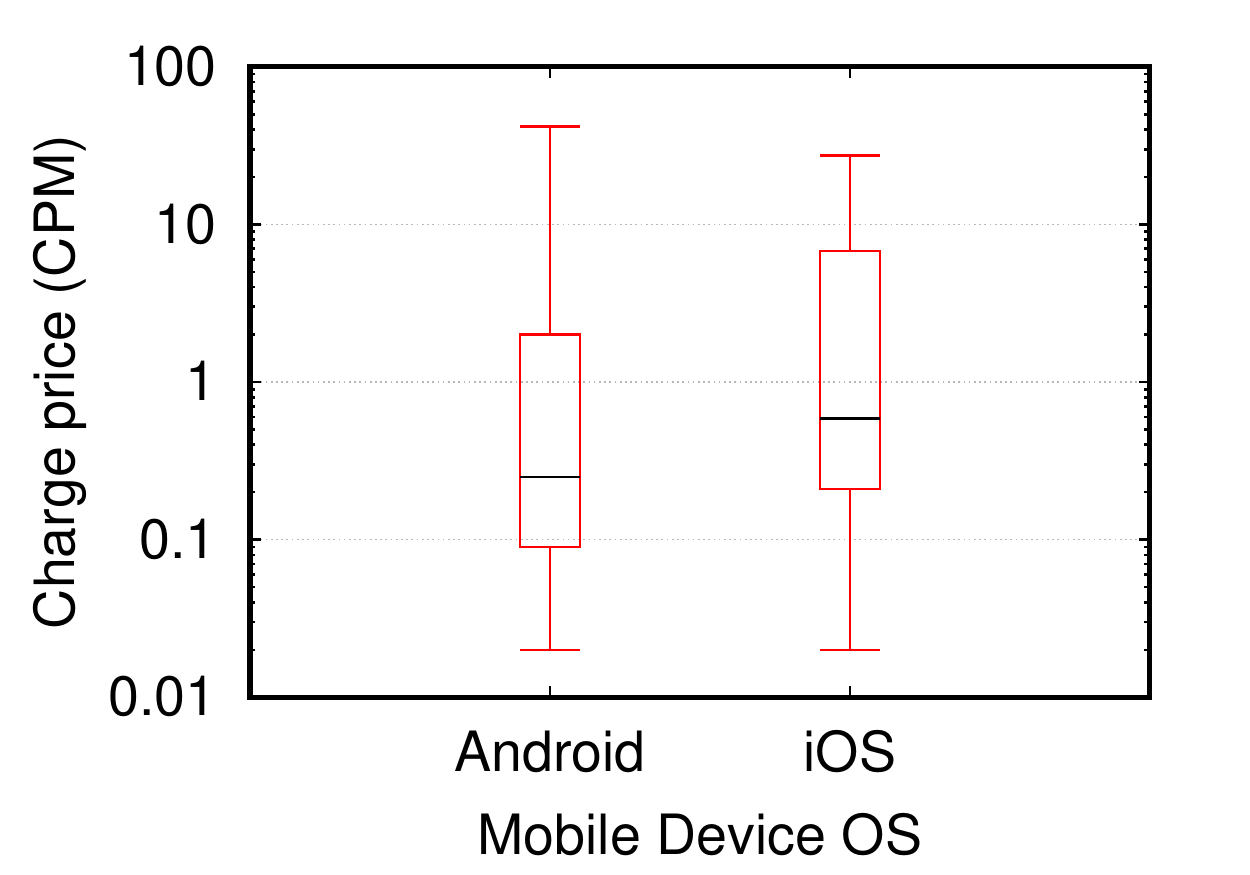}
		\caption{Distribution of charge prices per mobile OS.} 
		\label{fig:devicePricesDistribution}
	\end{minipage}
	\hfill
	\begin{minipage}{0.23\textwidth}
		\includegraphics[width=1.1\columnwidth]{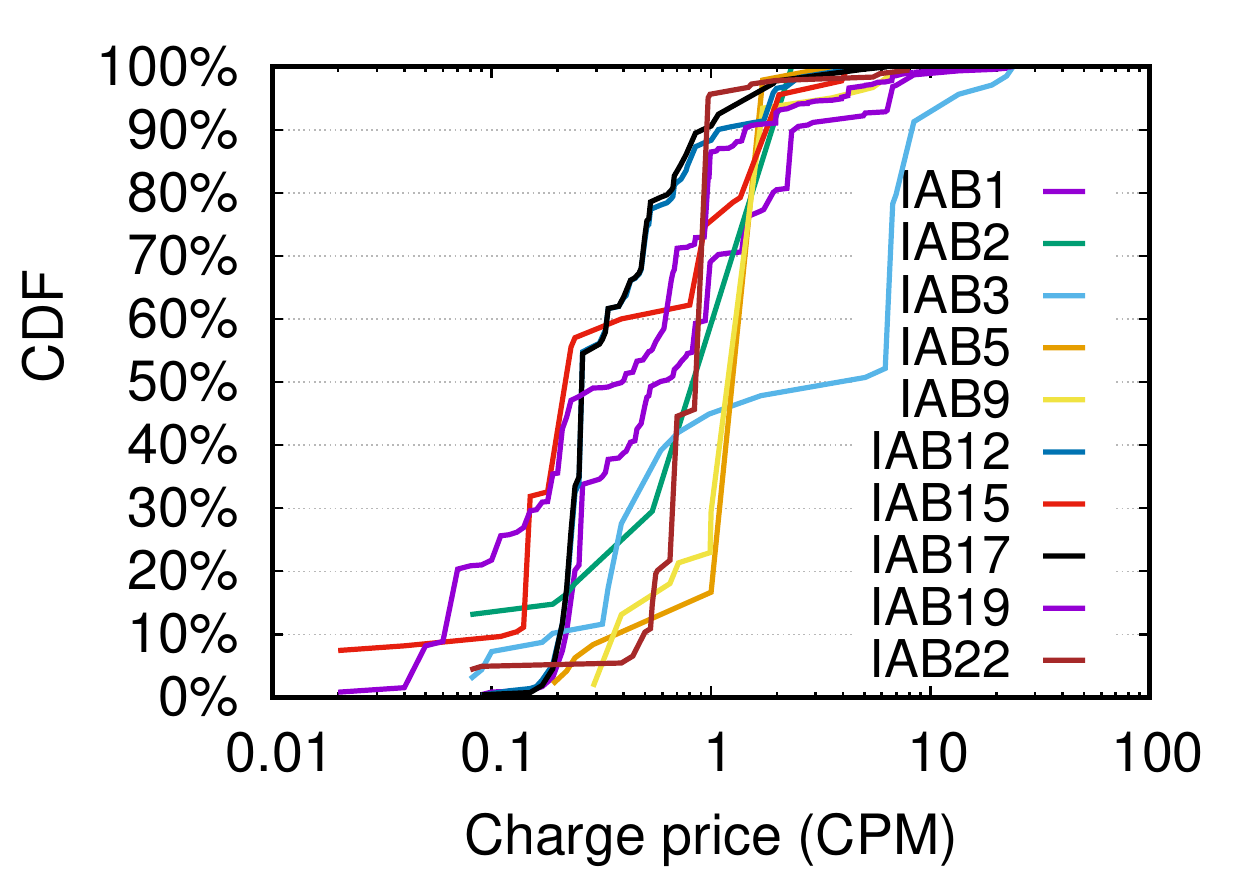}
		\caption{CDF of the generated cost per IAB category.} 
		\label{fig:iabCDF}
	\end{minipage}
\end{figure*}

\section{Bootstrapping PME}\label{sec:dataset}
We assess the feasibility and effectiveness of our methodology by bootstrapping the PME to train our model on real data by collecting a year long dataset containing weblogs from 1594 volunteering mobile users from the same country. 
Our users agreed to use a server of our control as a proxy, allowing us to monitor their outgoing HTTP traffic.\footnote{Data were treated anonymously although users signed a consent form allowing us to collect and analyze their data.}
As a result, we were able to collect a large dataset $D$ of 373M HTTP requests spanning the entire year of 2015.
Note that though our dataset consists of HTTP-only traffic, in principle our approach works with HTTPS as well, using as input the users' contributed data as can be seen in Figure~\ref{fig:system}.
Table~\ref{tbl:summaries} presents a summary of our dataset $D$.
Next, we present the data collection and analysis to extract features used in the price modeling and ad-campaign planning.

\subsection{Dataset analysis}
\noindent 
{\bf Weblog Ads Analyzer.}
To process our dataset, we implemented a weblog advertisements analyzer, capable of detecting and extracting RTB-related ad traffic.
First, the analyzer uses a traffic classification module to categorize HTTP requests based on an integrated blacklist of the popular browser adblocker Disconnect~\cite{disconnect}.\footnote{Our analyzer can also integrate more than one blacklists (e.g., Adblock Plus' Easylist, Ghostery's blacklist, etc.)}
Using this blacklist, the analyzer categorizes domains in 5 groups based on the content they deliver: (i) Advertising, (ii) Analytics, (iii) Social, (iv) 3rd party content, (v) Rest.
It consequently applies a second-level filtering on the advertising traffic by parsing each URL for any RTB-related parameters (like \nurl).
The analyzer detects \nurls\ by applying pattern matching against a list of macros we collected after 
(i) manual inspection and past papers~\cite{lukasz2015rtb2,lukasz2014selling-privacy-auction}, and
(ii) studying the existing RTB APIs~\cite{pulsepoint2016RTBmacros,doubleclick2016RTBmacros,openx2016RTBmacros,iab2015openRTB-2.4,mopub2016RTBmacros} 
used by the current dominant advertising companies.
From these detected~\nurls, it extracts the charge prices which we assume in this study that are in US dollars\footnote{Given that the majority of ADXs are located in US and following previous works~\cite{lukasz2014selling-privacy-auction}, we assume every charge price to be in US Dollars (so $1 CPM = $1/1000 impressions).} paid by the winning bidders, after filtering out any bidding prices that may co-exist in each \nurl.
It also extracts additional ad-related parameters such as ad impression ID, bidder's name, ad campaign ID, auction's ad-slot size, carrier, etc.

Other operations carried out by our analyzer include:
(i) user localization based on reverse IP geo-coding,
(ii) separation of mobile web browser and application originated traffic based on the {\tt user-agent} field of each HTTP request, 
(iii) extraction of device-related attributes from the {\tt user-agent} field (type of device, screen size, OS etc.), 
(iv) identification of cooperating ADXs - DSPs pairs, leveraging the \nurl\ used by the ADX to inform the bidder (i.e. DSP) about its auction win,
(v) user interest profile based on web browsing history.

\noindent 
{\bf Feature extraction.}
DSPs use different machine learning algorithms for their decision engines, taking various features as input, each affecting differently the bidding price and, consequently, the charge price of an ad-slot.
To identify such important parameters, we extracted several features from the \nurls\ of our dataset such as user mobility patterns, temporal features, user interests, device characteristics, ad-slot sizes, cookie synchronizations~\cite{Acar:2014:WNF:2660267.2660347}, publisher ranking, etc.
Next, we present the analysis of the most interesting features (Table~\ref{tab:aggregated-features} presents a summary).
We group them into 3 categories: geo-temporal state of the auction (\S~\ref{subsec:geo-temporal}), user's characteristics (\S~\ref{subsec:user-related}), and ad-related (\S~\ref{subsec:ad-related}).

\vspace{-0.2cm}
\subsection{Geo-temporal features}
\label{subsec:geo-temporal}

An important parameter that affects the price of an RTB ad is the user's current location~\cite{Guha:2010:CMO:1879141.1879152}, information which is broadly available to publishers and trackers.
Thus, in our dataset we extract user IP address and using the publicly accessible MaxMind geoIP database~\cite{maxmind}, we map each IP to its city level.
In Figure~\ref{fig:locationsWiskers}, which presents the 5th, 10th, 50th, 90th and 95th percentile of  the charge prices, we see that although the median values are relatively lower in large cities, the fluctuation of their price values is higher.

Another important feature is time, and specifically the time of day and day of week.
This is important due to the different level of attention a user may give to an ad impression and the amount of time she has to purchase an advertised product (e.g., working hours vs. afternoon's free time, or weekdays vs. weekends).
In Figure~\ref{fig:TODWiskers}, although the median charge prices are of similar range, we see that the early morning hours until noon tend to have more charge prices with increased values.
In Figure~\ref{fig:DOWWiskers}, we see a periodic phenomenon, where although in median values the charge prices are quite close, during weekdays the max prices are relatively higher than on weekends.\footnote{For time-of-day and day-of-week distributions, which visually appear to be similar, we confirmed that they are, in fact, statistically different with non-parametric, two-sample Kolmorogov-Smirnoff tests at p-value levels of $p_{tod}$$<$$0.0002$ and $p_{dow}$$<$$0.002$.}

\begin{figure}[t]
	\centering
	\includegraphics[width=0.85\columnwidth]{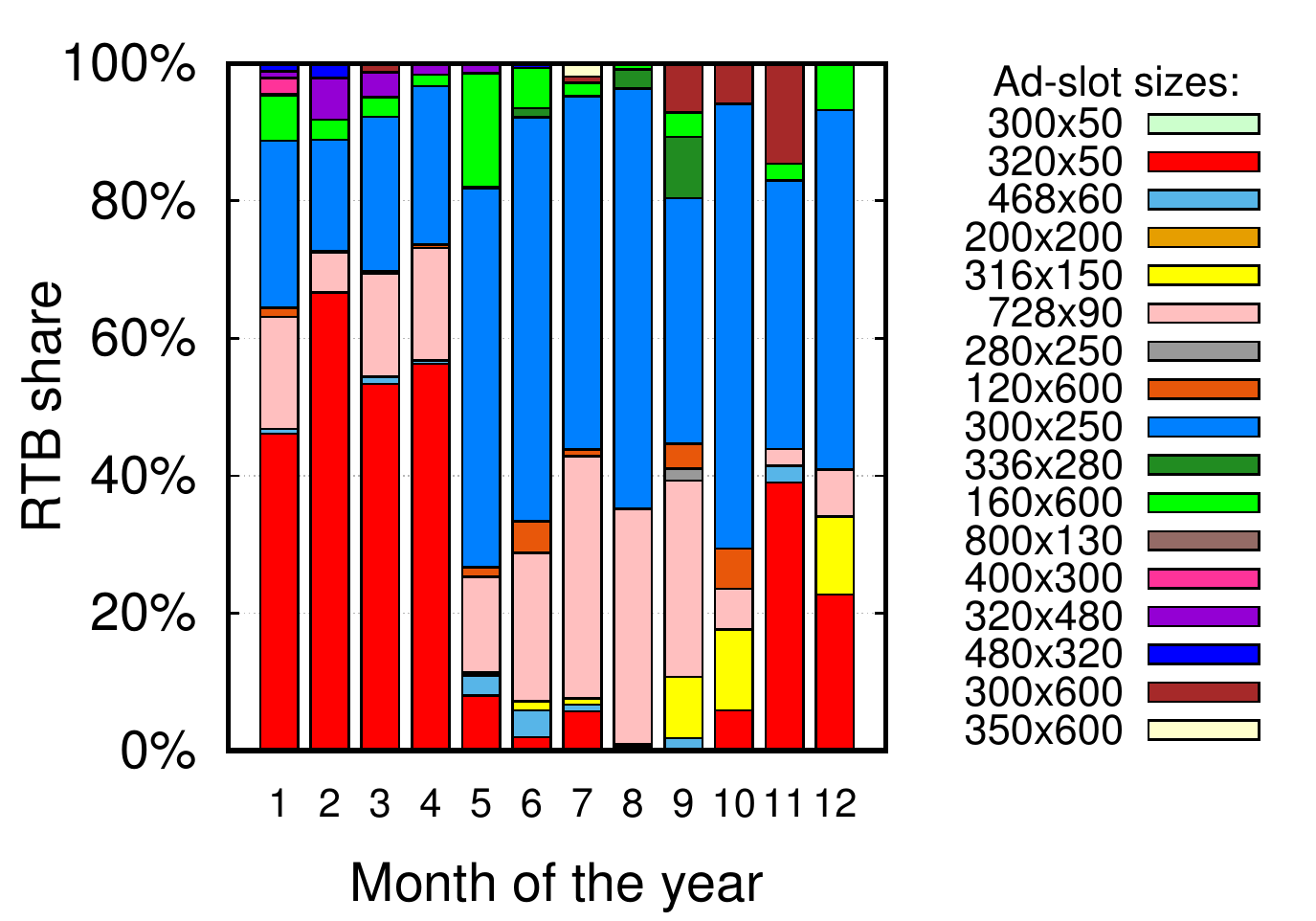} \vspace{-0.2cm}
	\caption{Ad-slot size popularity through time (sorted by area size).}
	\label{fig:adSizePopularity}
\end{figure}
\begin{figure}[t]
	\hfill
	\begin{minipage}{0.23\textwidth}
		\centering
		\includegraphics[width=1.05\columnwidth]{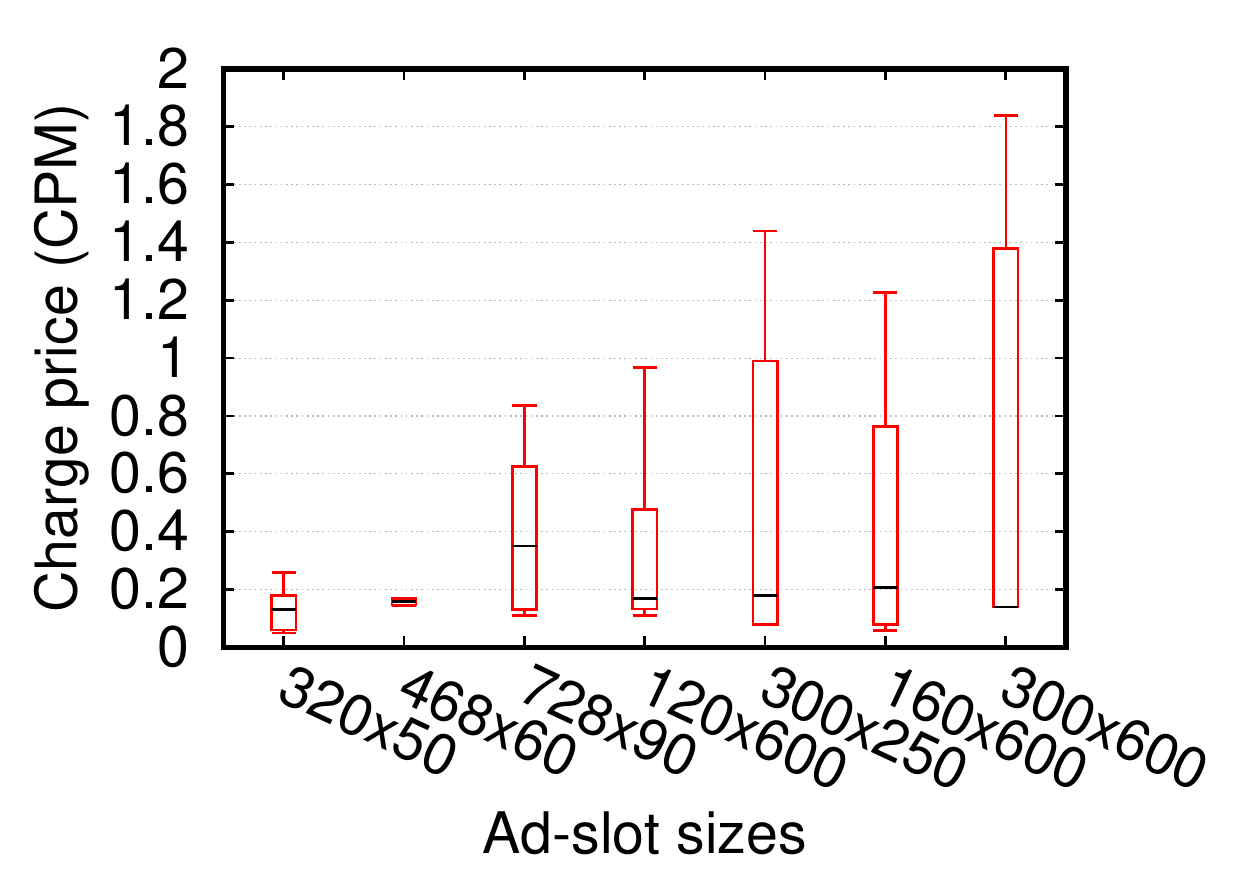}
		\caption{Distribution of the charge prices per ad-slot size (sorted by area size).}
		\label{fig:sizePricesDistribution}
	\end{minipage}
	\hfill
	\begin{minipage}{0.23\textwidth}
		\includegraphics[width=1.05\columnwidth]{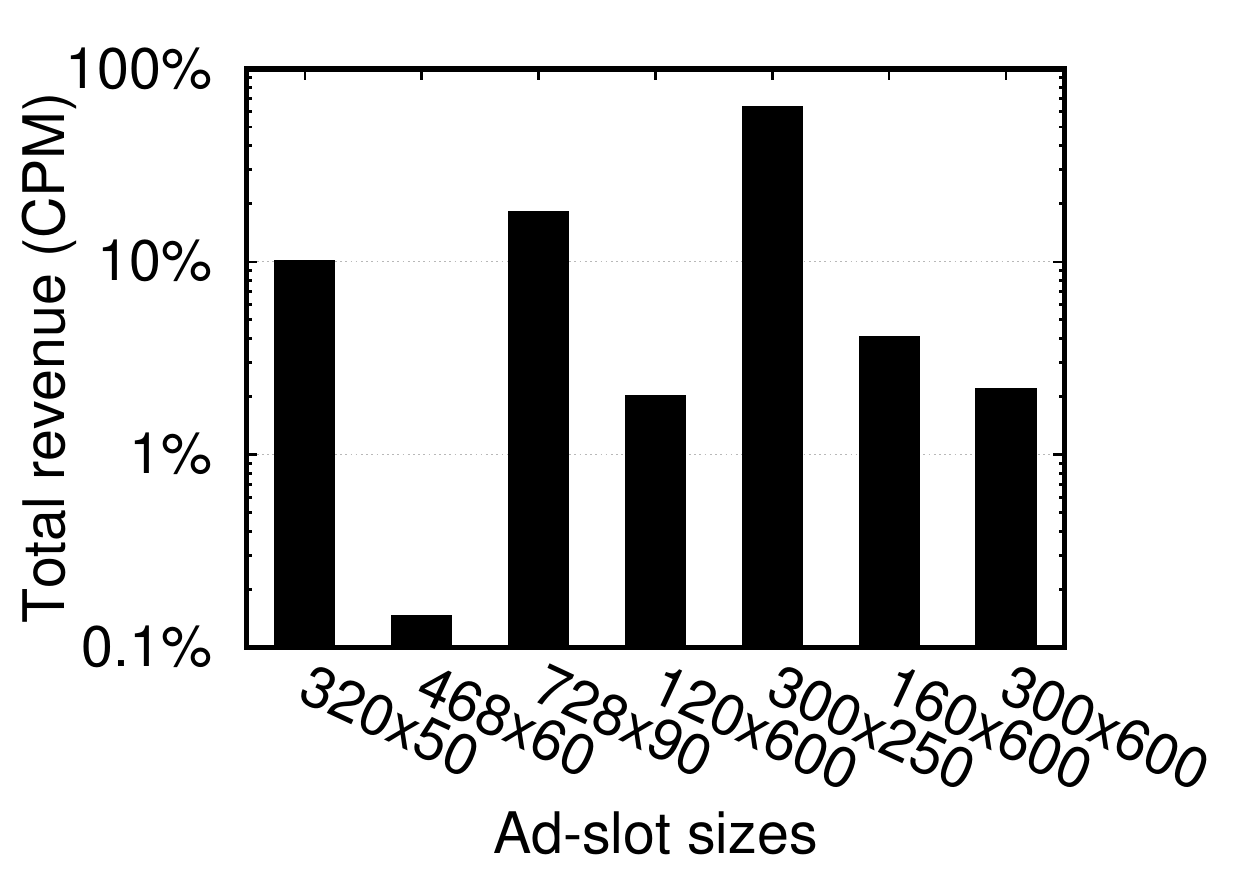}
		\caption{Accumulated revenue per ad-slot size (sorted by area size).}
		\label{fig:sizePricesTotal}
	\end{minipage}
\end{figure}

\subsection{User-related features}
\label{subsec:user-related}

\noindent 
{\bf Device type.}
By parsing the {\tt user-agent} (UA) header information, our analyzer classifies traffic and inspects the different fingerprints the UA leaks (specifications of process virtual machine (e.g., Dalvik or ART) or kernel (e.g., Darwin), operating system, browser vendor etc.)
Thus, we are able to identify the type of device (PC or mobile), the different types of mobile operating systems (Android, iOS, Windows) and if the traffic was generated from a mobile app or a mobile web browser.

In Figure~\ref{fig:devicePopularity}, we see the percentage of RTB traffic for the different OSes over time.
As expected, Android and iOS dominate, owning the larger portions of the market through the entire year, with Android-based devices appearing in 2x times more RTB auctions.
However, when normalizing this RTB share per mobile OS (Figure~\ref{fig:normPopularity}), we find that Android and iOS devices are delivered mostly equal RTB impressions, with some months Android surpassing iOS and vice-versa.
Then, we extract the traffic originated from the most popular ad-entity, MoPub~\cite{mopub}, and analyze the charge prices of the impressions rendered in the different OSes.
Surprisingly, although Android-based devices are more popular, we see in Figure~\ref{fig:devicePricesDistribution} that iOS-based devices tend to receive higher RTB prices, in median values.


\noindent 
{\bf Inference of the user's interest.}
The browsing history of a user is used by the advertising ecosystem as a proxy of her interests.
By monitoring the websites a user visits through time, a tracker can infer her interests, political or sexual preferences, hobbies, etc., quite accurately~\cite{bh-interests}.
To enrich our set of features with the users' interests, we collect all the websites each user visits across her whole network activity.
Such information is available to the RTB ecosystem as well, by using cookie synchronization~\cite{Acar:2014:WNF:2660267.2660347} or web beacons~\cite{webbeacons}.
To extract the interests from the visited websites, similar to existing approaches~\cite{Barford:2014:AHA:2566486.2567992}, we retrieve the associated categories of content for each website according to Google AdWords~\cite{adwords}.
Then, we aggregate across groups of categories for each user and get the final weighted group of interests for each user in the form of IAB categories~\cite{iabCats}.
Figure~\ref{fig:iabCDF} presents for the top mobile ADX (MoPub) a distribution of the generated ad revenue for the different IAB content categories in a 2 month subset of our dataset.
As expected, not all IAB categories cost the same.
Indeed, there are categories that are associated with products which attract higher bid prices in auctions, like IAB-3 (Business \& Marketing), with an average charge price of up to 5 CPM for the $50\%$ of the cases.
Alternatively, there are categories like IAB-15 (Science), which are unable to draw prices higher than $0.2$ CPM for the $50\%$ of the cases.


\subsection{Ad-related features}
\label{subsec:ad-related}

\noindent
{\bf Web Vs.~Apps}
Advertisers bid for ad-space in both webpages and mobile apps. After studying the cost per ad in both counterparts in our dataset, we see that apps draw on average 
2.6$\times$ higher prices (0.712 CPM vs. 0.273 CPM). This is expected; studies have shown that more advertising budget is spent on mobile application ads instead of 
mobile web, driving higher prices per ad~\cite{WebVSAppAds}: (i) Users pay more attention to app ads as they typically occupy fixed places in the screen, with no 
opportunity to scroll them out of sight as in web ads. In addition, (ii) studies~\cite{appsVsWeb} have shown that apps leak more personal data to advertisers, enabling them to deliver more targeted ads.

\noindent
{\bf Ad-slot sizes.}
Some ad-entities carry in their \nurls\ a parameter with the size of the auctioned ad-slot.
In Figure~\ref{fig:adSizePopularity} we plot the popularity of each of the ad-slot sizes through time.
It's interesting to see that 300x250 ad-slots (known as ``MPUs'' or ``Medium Rectangles'') dominate the dataset from May'15 on, taking the place of 320x50 ad-slots (known as ``large mobile banners'').
In fact, 300x250 ad-slots have more ad content available from advertisers, so they can increase earnings when both text and image ads are enabled.
In addition, we see that the 728x90 ad-slot (``leaderboard'' or ``banner'') is also popular.
This ad-slot, usually placed at the top of a page, is seen by users immediately upon page load.

It is easy to anticipate, that the more space an ad-slot covers in the user's display, the higher the price will be.
To verify this intuition, we isolated the traffic of an ad-entity (i.e. Turn~\cite{turn}), which carries the ad-slot size in its 
\nurls\ along with the associated charge prices.
Surprisingly, in Figure~\ref{fig:sizePricesDistribution}, we see that this intuition is wrong since the most expensive ad-slots for an advertiser are in fact, not the largest ones.
In our dataset, we see that the two most expensive ad-slots are the 
MPU (300x250) and Monster MPU (300x600), with median prices of 0.47 and 0.39 CPM, respectively.
However, from Figure~\ref{fig:sizePricesTotal}, the increased popularity of MPU and Leaderboard ad-slots, allows them to accumulate 64.3\% and 20.6\% of the total RTB revenue of Turn in our dataset, respectively.
Finally, it is worth to note that our results verify past resources~\cite{adSizePrices,adSizeGuide} regarding the more expensive ad-slots.

\subsection{Summary}
In summary, by analyzing the features extracted from our offline dataset, we find that a user's location (at city level) affects the median price that advertisers pay as well as its variability.
However, such price differences are expected to be more evident at the country-level, as shown in~\cite{lukasz2014selling-privacy-auction}.
In addition, the days and hours that a user may not be busy (Sundays), or may offer more attention (e.g., early mornings, Mondays) lead to higher charge prices.
The type of user's device also affects the charge prices but in a rather contradicting fashion: though there are more Android devices, iOS-based devices draw higher median prices.
As expected, the total revenue per category of user interest (through IABs) varies a lot, with some IABs being more costly than others.
Finally, the display's real-estate occupied by an ad-slot does not correlate well with price.
In fact, larger ad-slot sizes do not mean higher prices.
As shown in the next section, these extracted features are used to plan effective ad-campaigns and model encrypted charge prices.

\section{Charge Price Estimation}
\label{sec:price-modeling}

In order to create a model that estimates the encrypted prices detected on the user's browser and computes the total cost advertisers pay for her personal data, we need to have ground truth on charge encrypted prices.
However, such dataset is not easy to acquire.
One way to obtain this information is to collaborate directly with an ADX that sends such encrypted prices.\footnote{We considered top ADXs for encrypted prices (DoubleClick, OpenX, RubiconProject, PulsePoint), and ADXs for cleartext prices like MoPub (top mobile ADX).}
We assume this to be the rare case, since ADXs are generally unwilling to share such kind of data that may reveal bidding strategies and revenues.

In order to collect ground-truth data on encrypted prices, our system conducts small probing ad-campaigns on ADX(s)-DSP pairs that encrypt the winning prices. 
Such ad-campaigns can be designed and executed with the help of a single or few DSPs, with little overhead and a small budget of a few hundred dollars.
In addition, they can be optimized by using a specific set of experimental setups, which cover all possible scenarios from the small parameter vector $S$ to be kept short, efficient and cheap.
Given that the prices do not change drastically over time, these campaigns can be executed every few months to collect probing data for \emph{time-shift correction} and increased coverage of more ADXs. 
Besides, they can be automated and re-launched as frequently as needed, e.g., every few months or when the detected cleartext prices deviate from historical data.
Having such campaigns launched from a specific location allows for more accurate and cost efficient price modeling that can be shared across all participating platform users in the same area or country.

We envision that such campaigns can be crowd-funded (like Tor Project~\cite{tor}, Wikipedia, WiGLE~\cite{wigle}, etc.), thus, contributing to an independent and sustainable platform that can scale better across users, countries, and ADXs covered.
One may argue that these probing campaigns could pollute users' browsing with non-useful ad impressions.
Thus, they need to comply with the current standards, and if possible, consider an actual product or service.
Of course, ADXs could in principle fight back and try to identify and block such campaigns, but their huge clientele combined with the low volume of such campaigns makes the detection very difficult.
Next, we describe the effort to select a subset of core features important for price modeling (\S~\ref{sec:dimensionality-reduction}) and how they allow us to design efficient and effective ad-campaigns (\S~\ref{subsec:ad-campaigns-setup}).
Then, we provide an analysis of the data collected by two such campaigns (\S~\ref{subsec:ad-campaigns-data}) and we describe the model that estimates encrypted prices, which can be used by end-users (\S~\ref{subsec:encrypted-price-modeling}).

\subsection{Dimensionality reduction of features}
\label{sec:dimensionality-reduction}


The cost of testing all possible combinations of parameters and their values from the available feature set $F$ (with one probing ad-campaign each), would constitute the budget for the ad-campaigns impossible (1000s of setups x 10s Euros/setup).
Therefore, to perform ad-campaigns that are both effective and cost efficient, we need to select a subset of features $S\subseteq F$ that best describe the RTB prices found in weblogs such as the historic dataset $D$.
This subset of features should explain as much of the variability of prices as possible.
Assuming both encrypted and cleartext prices are affected by the same set of important features, this set should be small.
The fewer features we select as important, the smaller the cost of running ad-campaigns to collect representative RTB prices using these features (e.g., 10s-100s of setups).

To achieve this selection, we performed dimensionality reduction using all the available features (288) described in Section~\ref{sec:dataset} and Table~\ref{tab:aggregated-features}, using the cleartext prices as the target variable for optimization.
Some of these features are dense, i.e., they have an actual value in each price (e.g., time of day, day of week, size of ad, etc.) and others are sparse (e.g., interest categories of the user through time, publishers visited by user through time, etc.).
First, for normalization, we applied a log transformation on the extracted cleartext prices found in $D$.
Then, we applied a clustering of the prices into 4 classes, using an unsupervised equidistance model that finds the optimal splits between given prices using a method of leave-one-out estimate of the entropy of values in each class.
Next, we filtered out features that did not vary at all (i.e., constants) or had very high variance (99\%) (i.e., likely to be noise).

As a final step, dimensionality reduction (or feature selection) techniques such as PCA or Random Forests (RF) can be used~\cite{knime2015dimreduction}.
We chose the RF model\footnote{An ensemble of decision trees built using a random subset from the available features.} because it takes into account the target variable (cleartext price), it can be trained quickly on large datasets, it maintains interpretability of features and generally does not overfit the given data.
In case the availability of cleartext prices is limited, the reduction step to identify important features to be used in ad-campaigns could be hindered.
To mitigate this, the PME can use intermediate techniques such as high correlation filters that do not require a target variable, to eliminate features carrying similar information.

We trained various RF models using subsets of semantically related features from the available feature set and the best features from each subset were selected based on their power to describe the cleartext price distribution.
In summary, we grouped features in the following sets:
A) time,
B) http-related,
C) advertisement-related,
D) DSP-related,
E) publisher/host interests,
F) user http statistics (historical),
G) user interests (historical), and
H) user locations (historical).
We also tried selecting representative features out of each set to create minimal combinations covering all aspects of the http-available information.

In total, we tried tens of feature subsets and combinations and evaluated them 
using standard machine learning metrics such as precision, recall, weighted area under the receiver operating characteristic curve (AUCROC) and out-of-bag error.
Dimensionality reduction could, in principle, lead to loss of accuracy in the effort to explain price classes.
However, our experimentation lead to a small subset of features with minimal loss of precision ($<2\%$) and recall ($<6\%$).
In fact, we conclude that an optimal subset that performs very well and is small enough to allow cost efficient ad-campaigns is a set that \emph{combines features from different groups}.
In particular, also confirmed with an ad-campaign expert, we select the following features to be used for the probing ad-campaigns described next:
S=\{application/web-browsing, device type, user location, time of day, day of week, ad format (size), type of website, ad-exchange\}.

\begin{table}[t]
	\begin{small}
		\begin{tabular}{p{3cm}|p{4.8cm}}
			\toprule
			Filter name				&	Range of values (type)			\\
			\midrule
			Cities					&	Madrid, Barcelona, Valencia, Seville	\\
			Type of interaction		&	Mobile in-app, Mobile web			\\
			Time of day				&	12am-9am, 9am-6pm, 6pm-12am	\\
			Day of week				&	Weekday, Weekend				\\
			Type of device				&	Smartphone, Tablet				\\
			Type of OS		&	iOS, Android					\\
			Ad-format (smartphone)		&	320x50, 300x250, 320x480 or 480x320\\
			Ad-format (tablet)			&	728x90, 300x250, 768x1024 or 1024x768\\
			Ad-exchange				&	MoPub, OpenX, Rubicon, DoubleClick, PulsePoint\\
			Categories of targeting		&	all IABs possible				\\
			\bottomrule
		\end{tabular}
		\caption{Basic filters used in controlled ad-campaigns in Spain.
			In total, 144 experimental setups were attempted.}\vspace{-0.2cm}
		\label{tab:ad-campaign-filters}
	\end{small}
\end{table}

\subsection{Ad-campaigns setup}
\label{subsec:ad-campaigns-setup}

Using the most important parameters extracted in set $S$, we construct various experimental setups $s \in S \subseteq F$ that can be used to deploy such ad-campaigns over a short period of time $T'$ in selected ADXs to match top ADXs found in $D$.
These setups combine different values of control variables that are important for an ad-campaign:$<$user location, web-interaction type, time of day, day of week, device type, OS, ad-size, ADX$>$.
For example, an experimental setup could be this:
$<$Madrid, app, 12am-9am, weekday, smartphone, iOS, 320x50, MoPub$>$ (144 setups, Table~\ref{tab:ad-campaign-filters}).
Clearly, using more features would increase coverage of different types of ads, but also the campaigns' cost.
Instead, by running controlled ad-campaigns with a small feature set, we can receive ground truth data about encrypted prices, thereby allowing us to train a model for such prices, in a reasonable ad-campaign cost.

Campaigns with ADXs that deliver cleartext prices also allow us to compare prices in different times and compute shifts in the price distribution due to time passed between the collection of dataset $D$ and present time.
To compensate for the loss of information from cleartext prices becoming less abundant, additional features available in professional ad-campaign planners (as in FDVT~\cite{fdvt}) could be used in the future to enhance the setups tested.
With the results of these campaigns (in essence, charge prices for RTB ads that fulfil a given setup $s$), the PME can train a model to estimate the cost of new ads with a given setup $s'$ close, or equal, to one tested, i.e. $s' \sim s \in S$.

\noindent
{\bf Number of required ad-campaigns.}
An important decision in running probing campaigns is how many of them to launch, and with how many impressions in each one, in order to obtain a good approximation of the underlying distribution of prices.
For this, we analyzed the ad-campaigns found for MoPub in $D$.
We identified 280 such campaigns in 2015, with mean and standard deviation of charge price of $m=1.84$ and $std=2.15$ CPM, respectively.
We use the process described in~\cite{samplesize2} and the next formulation to compute $d$, the expected error on the mean, assuming a suggested number of setups $n$, and ignoring the finite population correction adjustment (thus assuming a more conservative approximation of $n$)
$
d=\frac{Z_{\alpha/2} \times std}{\sqrt{n}},
$
where Z is the z-score of normal distribution.
Using the 144 setups proposed, we can approximate to more than 95\% CI (i.e., $\alpha$$=$$0.05$) the mean price of campaigns observed in the wild, assuming a margin of error 0.35 CPM.
Also, considering the distribution of prices within the largest of ad-campaigns detected for MoPub with 1.8k impressions, we can approximate to 95\% CI the mean price of a campaign, assuming an error 0.1 CPM and minimum of 185 impressions per campaign.

\subsection{Ad-campaigns analysis}
\label{subsec:ad-campaigns-data}

Using the above as guideline, we executed two rounds of different ad-campaigns to collect data on prices (Table~\ref{tbl:summaries}).
Our ad-campaigns advertised a real non-for-profit NGO in the area of data transparency, in an attempt to avoid polluting users with meaningless impressions, and trying to do something useful with the allocated budget.

\noindent
{\bf Dataset collected.}
The first round ($A1$) was executed for 2 weeks in May 2016 and utilized the 4 ADXs mentioned earlier (also found in $D$) that encrypt price notifications and targeted publishers of many IAB categories.
The second round ($A2$) was executed with the same experimental setups as $A1$ during June 2016, but in this case the DSP was instructed to use only MoPub, while still targeting similar IAB categories of publishers.
These constraints allowed us to directly compare encrypted with cleartext prices in the same period, and time-shift all prices detected in $D$ from 2015 to 2016.

In both campaigns, the DSP was given an upper bound on the bidding CPM price to safeguard that the allocated budget will not be consumed quickly.
Because studying the effects of retargeting is beyond the scope of this paper, we did not ask the DSP to perform such campaigns.
However, the DSP was instructed to bid in a dynamic manner, as low or high as needed to get the minimum of impressions delivered for the various experimental setups we requested.
We plan to investigate the effects of retargeting in a separate and dedicated future study.
Overall, we managed to receive across all setups, over $600k$ impressions displayed with encrypted price notifications to more than 200 publishers, and over $300k$ impressions with cleartext price notifications to more than 300 publishers, reaching audiences of 6 IAB categories common to both notification types.

\begin{figure}
	\centering
	\includegraphics[width=0.85\columnwidth]{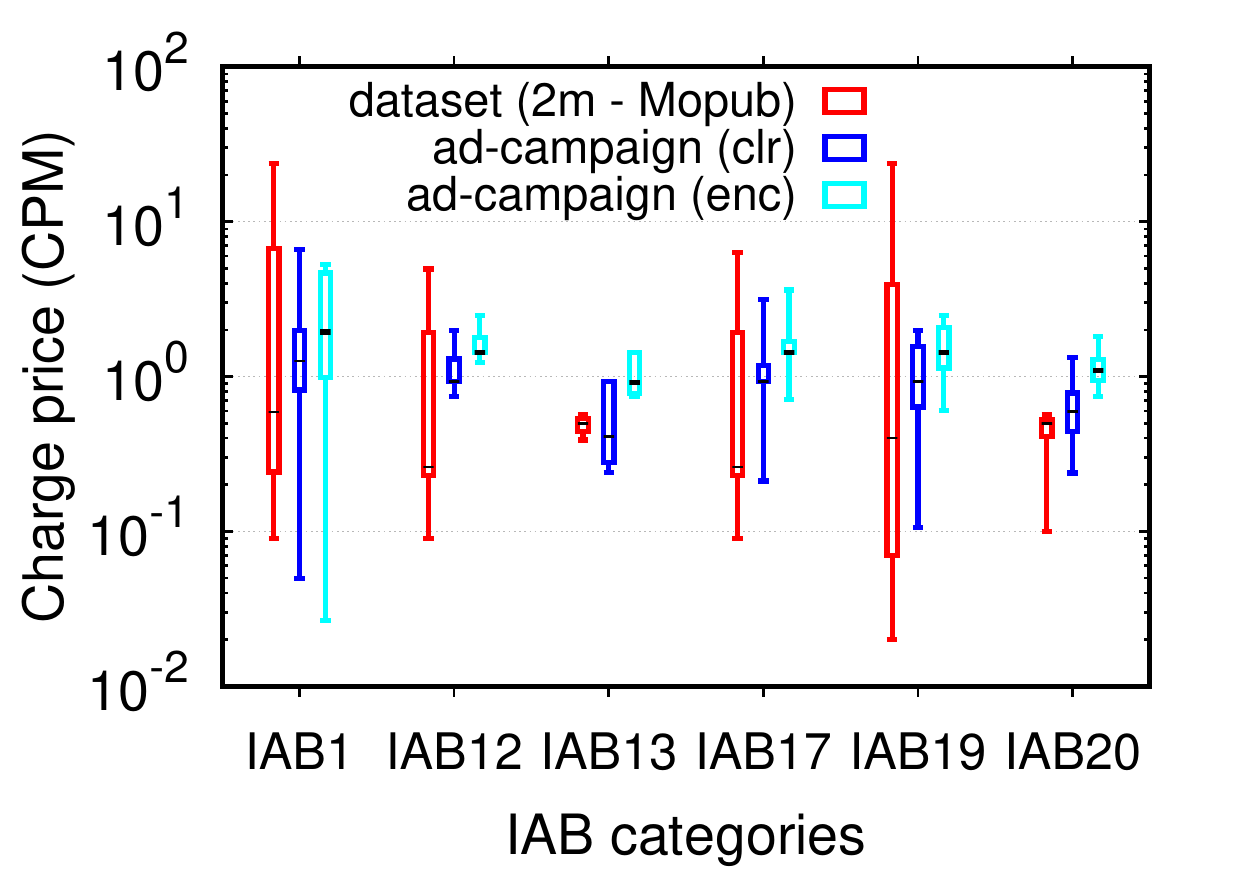}
	\caption{Comparison of CPM costs for the different IAB categories in our dataset and the 2 probing ad-campaigns.}
	\label{fig:iabBars}
\end{figure}

\noindent
{\bf Cost paid vs. IAB category.}
In Figure~\ref{fig:iabBars}, we compare the overlapping IAB categories of the RTB impressions we took from (i) the set of encrypted prices from the ad-campaign on four ADXs in $A1$, (ii) the set of cleartext prices from the ad-campaign on MoPub ($A2$), (iii) the 2 months MoPub subset of $D$.
Note that in some cases, the results from $D$ vary more than in the ad-campaigns.
This is to be expected, as the dataset includes prices from numerous DSP-ADX pairs for many ad-campaigns running in parallel in the duration of a year, whereas our two ad-campaigns are more targeted to specific DSP-ADX pairs.

Regarding the cleartext prices of different IAB categories, although the median prices are usually in the same order of magnitude, they are higher in the case of the recent ad-campaign contrary to the 2 month dataset.
We believe that this difference is due to the time shift between the dataset collected in 2015 and the ad campaigns performed in 2016.
In addition, we see that the median price is always higher in case of encrypted prices ($A1$), compared to the cleartext prices of the second ad-campaign ($A2$) and dataset $D$.

\subsection{Encrypted price modeling}
\label{subsec:encrypted-price-modeling}

Using the ground truth data collected from the first round of ad-campaigns (encrypted prices) with various parameters within the subset of features $S$, we trained a machine learning classifier to predict values of encrypted prices.
We note that given the problem of modeling real values, we first applied regression models with different combinations of dependent variables ($S$).
However, the high variability of charge prices lead to low performance (high error) of the regression models.
Therefore, we proceeded to split the prices into groups for classification.
As a first step, we performed similar preprocessing for the encrypted prices as we did earlier for the cleartext prices (normalization and clustering to 4 classes of well balanced groups).
Next, we trained a RF model to predict the class of an encrypted price, based on the available parameters $S$.
For the training and testing, we applied 10-fold cross validation, and averaged results over 10 runs.
Using features such as city of user, day of week and the time the ad was delivered, ad size, mobile OS of the user's device, IAB category of the publisher, ADX used and device type, our classifier can achieve a very good performance: $TP$$=$$82.9\%$, $FP$$=$$6.8\%$, Precision$=$$83.5\%$, Recall$=$$82.9\%$, $0.964$ AUCROC.
These scores are weighted averages across all classes, with no class performing worse than 5\% from the average.
We repeated this process with more price classes (i.e., 5-10 groups) for higher granularity of price prediction, but the results with 4 classes outperformed them.

When the exact publisher used is also taken into account in the model, the performance of the classifier increases to $95\%$ accuracy, and $0.99$ AUCROC.
However, this is classic over-fitting and we should caution that the publishers used in the ad-campaigns are just a subset of the thousands of possible publishers that can be found in real weblogs.
Therefore, we chose to use the model with the IAB category but without the exact publisher as part of its input features.
Next, this model was used for the estimation of the encrypted prices of \nurls\ found in the weblogs of each user in $D$, given the matching parameter values from $S \subseteq F$.


\section{User Cost for Advertisers}
\label{sec:user-cost}

The previous sections allowed us to:
(1) bootstrap our price modeling engine from existing user weblogs, so that we find the important features describing well the observed RTB cleartext prices,
(2) using these important features, run probing ad-campaigns with ADXs that send encrypted price notifications, so that we collect ground truth on such prices from performance reports delivered to us,
(3) using such ground truth, train a machine learning model to estimate the price of new RTB notifications sent in encrypted form.
We are now ready to study the overall cost advertisers paid for each of the users in our dataset $D$, who received cleartext and/or encrypted prices in \nurls\ of delivered ads.


\subsection{Encrypted vs. cleartext price distributions}
\begin{figure}[t]
	\centering
		\centering
		\includegraphics[width=.8\columnwidth]{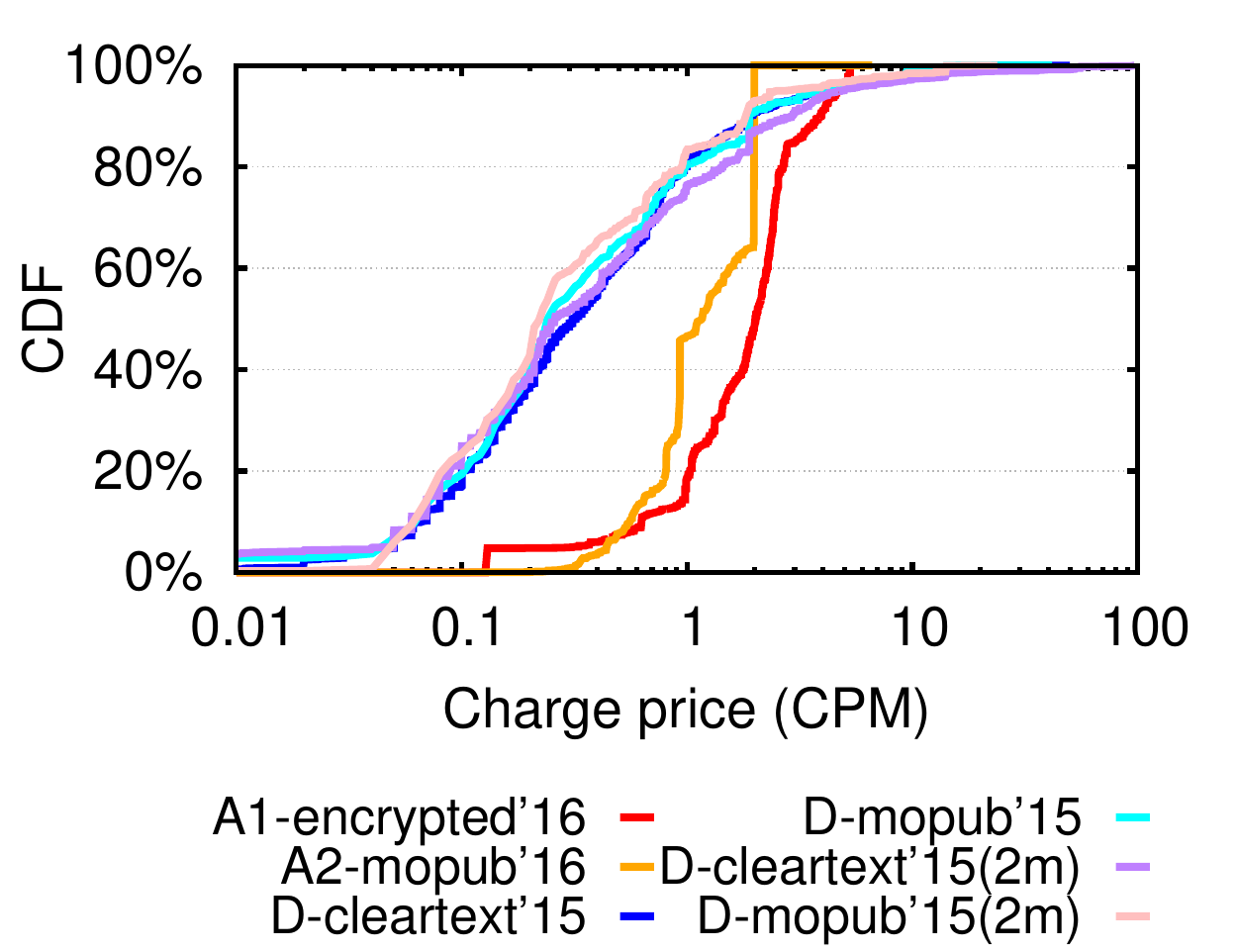}
		\caption{Comparison of price distributions between cleartext and encrypted, for different time periods and datasets ($D$~vs.~$A1$ and $A2$).}
		\label{fig:prices-comparison}
\end{figure}

The work in~\cite{lukasz2014selling-privacy-auction} assumed that encrypted prices follow the same distribution with cleartext prices.
To examine the validity of such assumption, we plot the distributions of both encrypted and cleartext charge prices we got from the two ad-campaigns we performed.
Interestingly, from Figure~\ref{fig:prices-comparison}, the distribution of {\it encrypted prices} in $A1$ is distinctly different and of {\it higher median value ($\sim$$1.7$$\times$) than cleartext prices} of $A2$.

In addition, we study the distributions between different time periods and ADXs to extract important lessons.
First, we see that the cleartext price distribution of MoPub (2015) is similar to all ADXs sending cleartext prices, either when considering a 2 month period or a full year.
Hence, we can study MoPub as a representative example and extrapolate lessons for the rest of the ADXs that send cleartext prices.
Second, the distribution of cleartext prices from $A2$ (MoPub) are of higher median value and can be used to establish the price shift due to time difference between the time $T$ the dataset was collected, and $T'$ when the campaigns were executed.
In reality, this price shift can be detected evenly across multiple probing ad-campaigns (e.g., once per quarter of year). 

\vspace{-0.1cm}
\subsection{How much do advertisers pay \\to reach a user?}
Equipped with our presented methodology for estimating encrypted prices, we are now ready to respond to our motivating question.
Specifically, we utilize our method and compute the overall cost advertisers paid for each user in the dataset $D$, i.e., across a whole year of mobile web transactions.
We also apply a time-correction coefficient on the cleartext prices using the prices from the second round of ad-campaigns.
This allows us to consider the increase in cleartext prices due to time difference from the weblog collection (2015) and the ad-campaigns execution (2016).

\begin{figure}[t]
		\centering
		\includegraphics[width=.8\columnwidth]{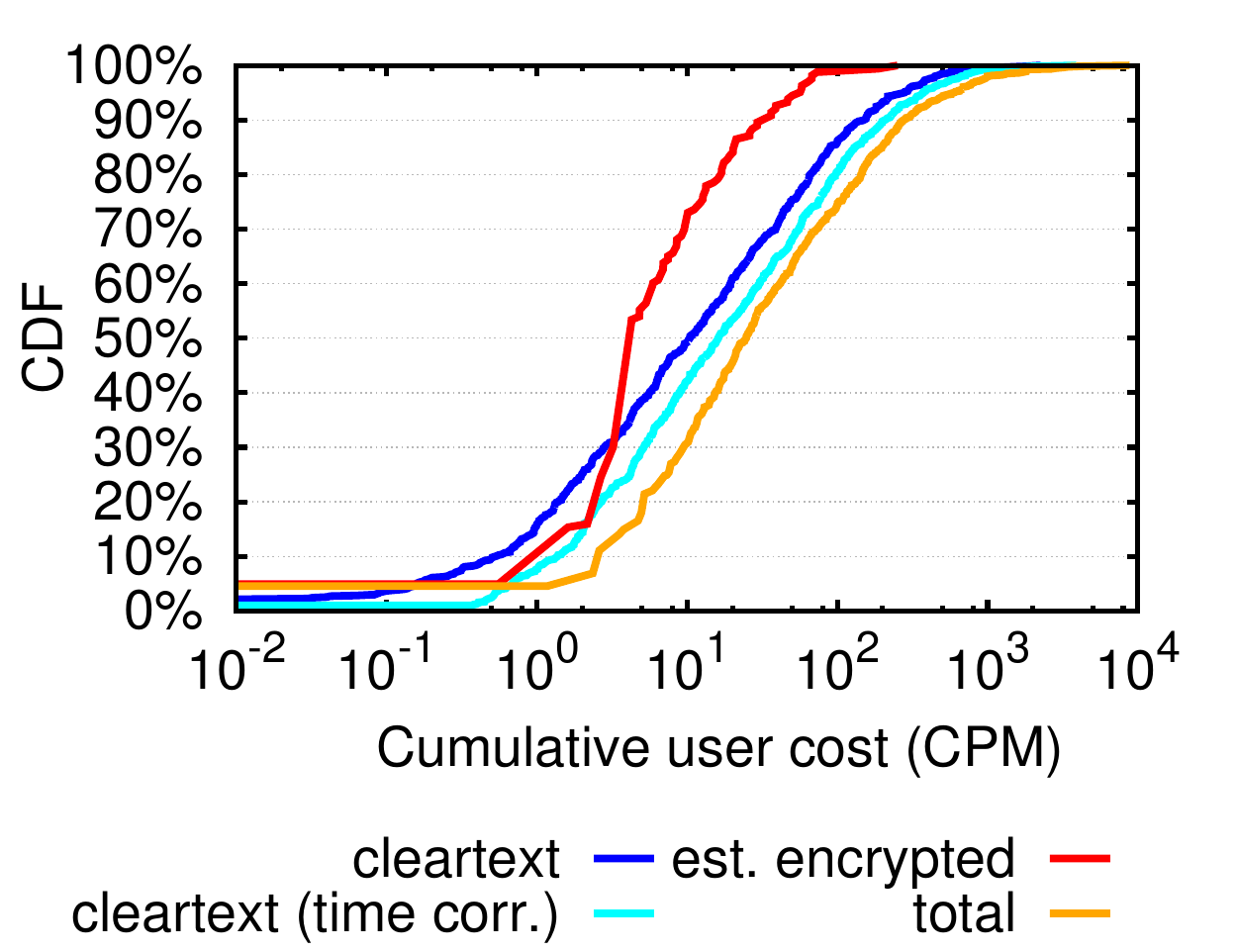}
		\caption{Cumulative CPM paid per user in our year long dataset.}
		\label{fig:totalUserCost}
\end{figure}

Figure~\ref{fig:totalUserCost} presents these cumulative costs in the form of CDFs of the price distributions.
As expected, we observe that the cumulative cost due to encrypted prices is still not surpassing the cleartext, since the latter is still the dominant price delivery mechanism in mobile RTB.
We also note that some users are more costly than others.
Specifically, the median user costs $\sim$$25$ CPM, and up to $73\%$ of the users cost $<100$ CPM through the year for the mobile ad ecosystem in the given dataset.
This means that the ad-ecosystem reaches the average user very cheaply and multiple times below what users estimate this cost to be (e.g.,10s of dollars~\cite{bigmac}).

On the other hand, for $\sim2\%$ of users, the advertising ecosystem spent $1000$-$10000$ CPM for the same time period.
Finally, about $60\%$ of users had an increased average cumulative cost of $\sim55\%$ on top of their cleartext cost, due to the estimated encrypted prices.
These users had a median of $14.3$ CPM added to their total cost, with some extreme cases of 1000-5000 CPM.

\begin{figure}[t]
	\centering
	\includegraphics[width=0.74\columnwidth]{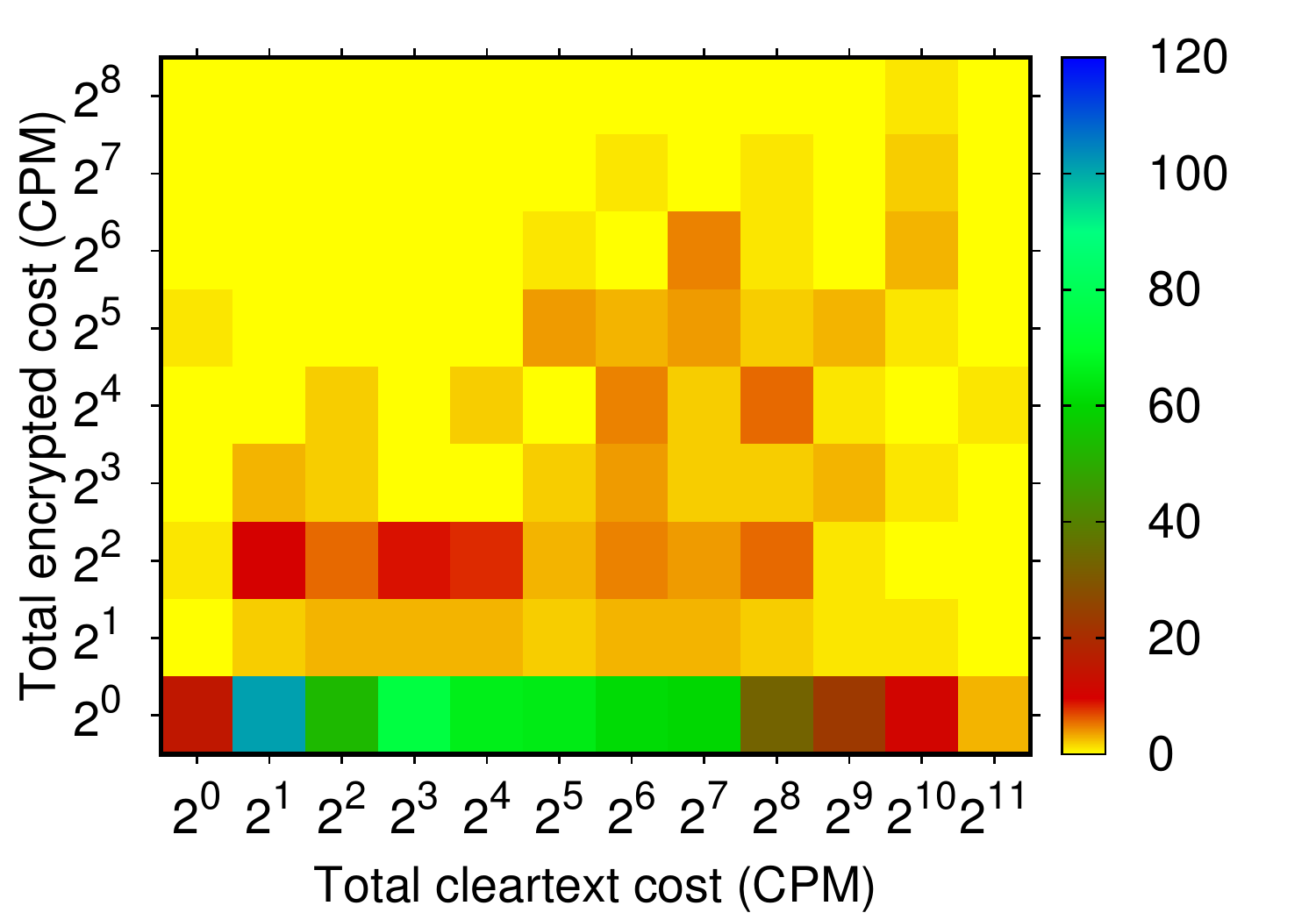}\vspace{-0.1cm}
	\caption{Total cleartext vs. total estimated encrypted cost of each user in $D$ (color indicates number of users).}
	\label{fig:clr_to_enc}
\end{figure}

\begin{figure}[t]
	\centering
	\includegraphics[width=0.8\columnwidth]{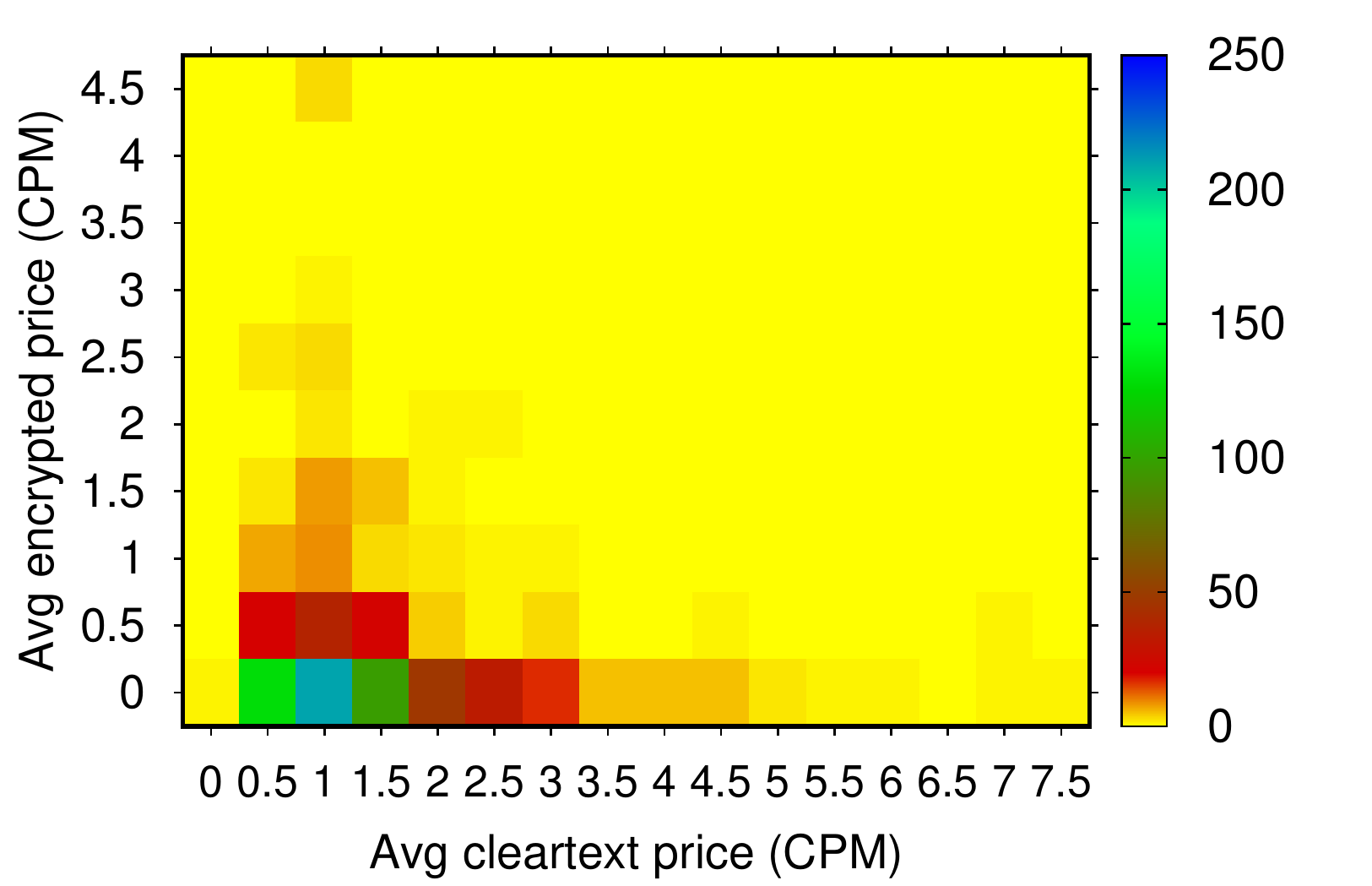}\vspace{-0.15cm}
	\caption{Average cleartext vs. average estimated encrypted price per impression of each user in $D$.}\vspace{-0.08cm}
	\label{fig:perImpr_clr_to_enc}
\end{figure}

In the previous result, we compared the distributions of encrypted and cleartext prices, while disregarding the targeted user.
In order to identify if the cost paid through encrypted prices is the same with cleartext for a specific user, we compare for each user the total costs in Figure~\ref{fig:clr_to_enc} and average cost per impression in Figure~\ref{fig:perImpr_clr_to_enc}, for each type of price.
We observe that a significant portion of users ($\sim$$20$-$25\%$) cost similarly for ads embedded with encrypted or cleartext prices.
As expected, due to the current majority of cleartext prices in the mobile ad market, a large portion of users ($\sim$$75\%$) have higher cumulative cost from cleartext than encrypted prices.
However, a small portion ($\sim$$2\%$) costs more ($2$-$32\times$) in encrypted than in cleartext form, because they were delivered mostly ads with encrypted prices.
When we normalize the cumulative ad cost of user per impression delivered (Figure~\ref{fig:perImpr_clr_to_enc}), we find that for small prices of $\leq$3 CPM/impression, cleartext is more dominant across users.
We also find a small portion ($\sim$$2\%$) of users who cost up to $5\times$ more CPM/impression for the delivered ads in encrypted than in cleartext form.
We anticipate this portion to increase as the encrypted notification becomes the dominant delivery of RTB prices in mobile.

\vspace{-0.2cm}
\subsection{Summary}
\label{sec:summary}
By studying the overall RTB advertising cost for users in our dataset, and distinguishing the encrypted from the cleartext prices, we found that the basic assumption of related work~\cite{lukasz2014selling-privacy-auction} that encrypted and cleartext prices are similar, is not valid (encrypted prices are around $1.7\times$ higher).
Furthermore, advertisers, based on users' personal data, paid $\sim$$25$ CPM for delivering ads to an average user, and less than $\sim$$100$ CPM for delivering ads to $3/4$ of users during a year.
We also identified a small portion of outlier users ($\sim$$2\%$) who cost $10$-$100\times$ more to the ad-ecosystem than the average user, and a similar portion that costs up to $32\times$ more in encrypted than cleartext prices, even though encrypted prices are only a quarter of the mobile RTB ecosystem.

\noindent
{\bf Validation.}
As an effort to validate our methodology, we can extrapolate how much users cost for the ad-ecosystem and if this estimation compares with current market numbers.
For this extrapolation, we make some assumptions on how our dataset represents the overall ecosystem of users and advertisers.
In particular, we assume that our average mobile user, whose annual ad-cost is in the $8$-$102$ CPM range (25th-75th perc.), has:
(1) performed $2.65$ hours online daily, which is $\sim$$83\%$ of the average daily mobile internet usage, when considering average tablet and other mobile device usage~\cite{smartphonevstablet},
(2) performed internet activity from both mobile and laptop/desktop devices, the former traffic type being $\sim$$51\%$ of total internet time~\cite{mobilevsdesktop},
(3) received ads in a similar fashion in both HTTP and HTTPS, the former being $\sim$$40\%$ of the total traffic delivered to a user~\cite{sandvine70https,httsUse},
(4) received ads over RTB, which has an overhead management and intermediaries cost of $\sim$$55\%$~\cite{rtboverhead}, and
(5) received ads in a similar fashion over RTB and traditional and other online advertising, the former being $\sim$$20\%$ of the total online advertising~\cite{rtbvsothers}.
Considering these factors, the overall average user ad-cost (25th-75th perc.) would be in the range of $\$0.54$-$6.85$, which is in the order of magnitude reported by major online advertising platforms such as Twitter (owner of MoPub, ARPU: $\$7$-$8$~\cite{twitterARPU}) and Facebook (ARPU: $\$14$-$17$~\cite{facebookARPU}) during the period 2015-2016. 



\section{Related Work}\label{sec:related-work}

There is a plethora of papers studying privacy loss and tracking techniques in the wild \cite{Krishnamurthy:2009:PDW:1526709.1526782, Acar:2014:WNF:2660267.2660347,englehardtonline,Nikiforakis:2013:CME:2497621.2498133,Eckersley:2010:UYW:1881151.1881152,Roesner:2012:DDA:2228298.2228315,Leung:2016:YUA:2987443.2987456}. 
There are also others proposing privacy preserving countermeasures based on either (i) randomization-/obfuscation- based techniques~\cite{ksub,Nikiforakis:2015:PDF:2736277.2741090}, where the authors aim to pollute the information trackers retrieve in order to hide the users' data and interests, or (ii) anti-tracking mechanisms~\cite{appsVsWeb,Kontaxis:2012:PSP:2362793.2362823}, where requests to trackers are avoided or blocked. 
All the above studies, highlight the voracity of web entities to collect data about the user and her online behavior, and an arms race between the privacy-aware users and trackers.

But how do all these trackers monetize from these data?
The answer is in the advertising ecosystem, where advertisers are purchasing audiences to deliver their ad-impressions.
Therefore, there are studies focusing solely on privacy preservation in the advertising ecosystem.
For example, Privad~\cite{Guha:2011:PPP:1972457.1972475} is designed to conceal user activities from an ad-network, by interposing an anonymizing proxy between the browser and the ad-network, allowing a trusted client software to select relevant ads locally.
Unfortunately, it requires broad adoption of high-performance anonymizing proxies.

Alternatively, Adnostic~\cite{toubiana2010adnostic} is an architecture for interest-targeted advertising without tracking.
Like Privad, Adnostic uses client-based functionality to perform ad selection, but eliminates anonymizing proxies at the cost of less precise ad targeting.
In~\cite{papaodyssefs2015wit}, authors propose obfuscation of the user's full identity while browsing the web.
This was achieved by introducing Web Identity Translator (WIT) in-between the user's client and the visited websites.
Given that advertisers are interested in adjusting their buying strategy at real time, it is unclear if such approaches can be adapted to contemporary technologies such as RTB auctions.

The economics of private data have long been an interesting topic and attracted a considerable body of research either from the user's perspective~\cite{acquisti2013privacy,bigmac,staiano2014moneywalks,forSale}, or the advertiser's perspective~\cite{lukasz2014selling-privacy-auction,followTheMoney,fdvt,yourdataworth}.
In~\cite{acquisti2013privacy} authors discuss the value of privacy after defining two concepts (i)~\emph{Willingness To Pay}: the monetary amount users 
are willing to pay  to  protect  their  privacy, and (ii)~\emph{Willingness  To  Accept}: the compensation that users are willing to accept for their privacy loss.
In two user-studies~\cite{bigmac,staiano2014moneywalks} authors measure how much users value their own offline and online personal data, and consequently how much they would sell them to advertisers. 
In~\cite{forSale}, the authors propose ``transactional'' privacy to allow users to decide what personal information can be released and receive compensation from selling them.

In~\cite{lukasz2014selling-privacy-auction}, the authors perform an analysis of cookie matching in association with the RTB advertising.
Similar to our approach, they leverage the RTB \nurl\ to observe the charge prices and they conduct a basic study to  provide some insights into these prices, by analyzing different user profiles and visiting contexts.
Their results confirm that when the users' browsing histories are leaked, the charge prices tend to be increased.
Similarly, in~\cite{olejnikbid}, the authors propose a transparency enhancing tool showing to the users the RTB charge price every time a RTB auction is performed.
Furthermore, they collect profiled and un-profiled data from a browser extension and a crawler respectively, and they compare the RTB prices, the bidding frequency and the inter-relations among ADXs and DSPs.
Contrary to our work, both studies use a dataset from (i) a small number of 100 users, (ii) over desktop, (iii) covering only one month, (iv) and based on these data, they estimate the advertising total revenues using only the cleartext prices based on the arbitrary assumption that encrypted and cleartext prices follow the same distributions.
Although their results regarding the average prices per ad are comparable to ours ($\sim$0.5CPM Vs. $\sim$0.26CPM), they are not equal since their study was conducted on desktop and in 2013, when ad spending in desktop was higher than in mobile~\cite{mobilevsdesktop}.

\begin{figure*}[t]
	\centering
	\includegraphics[width=1.255\columnwidth]{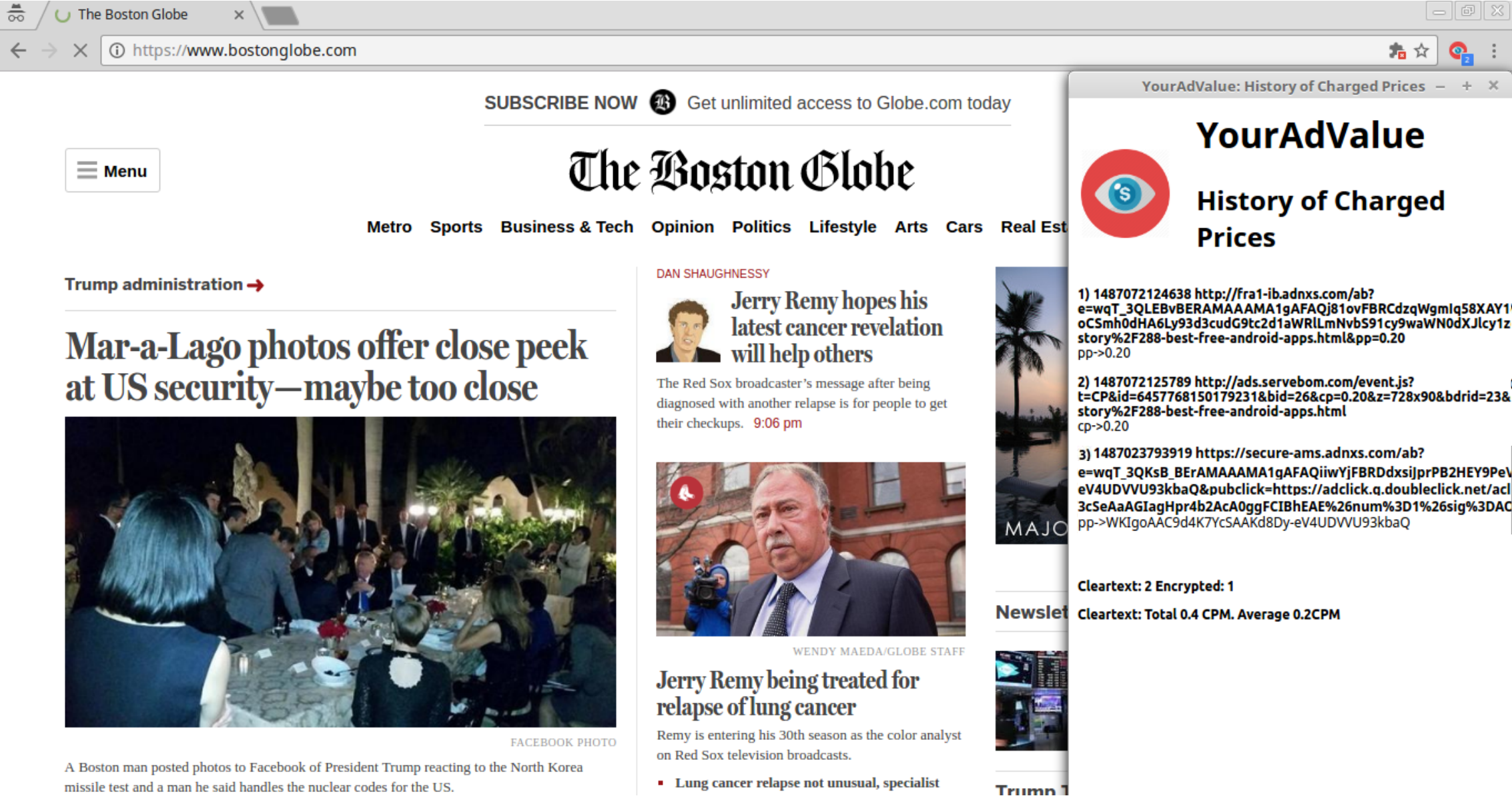}
	\caption{Preliminary implementation of \toolname\ Chrome extension in use.}
	\label{fig:pluginScreen}
\end{figure*}

In~\cite{followTheMoney}, authors use a dataset of users' HTTP traces and provide rough estimates of the relative value of users by leveraging the suggested bid amounts for the visited websites, based on categories provided by the Google AdWords. 
FDTV~\cite{fdvt} is a plugin to inform users in real-time about the economic value of the personal information associated to their Facebook activity.
Although similar to ours, our approach works for all HTTP activity of mobile users.
Furthermore, journalists from Financial Times, created an interactive calculator~\cite{yourdataworth} to explore how valuable specific pieces of user data are for the ad-companies.
This calculator is based on the analysis of industry pricing data from a range of sources in the US. 

Finally, the rapid growth of RTB auctions has drawn the attention of the research community, which aims to explore the economics of the RTB ad ecosystem.
In~\cite{yuanRTB}, the authors provide an insight to pricing and an empirical analysis of the technologies involved.
They use internal data of an ADX and they study its bidding behaviors and strategies.
In~\cite{RTBpredicting}, the authors propose a winning price predicting mechanism by leveraging machine learning and statistical methods to train a model using the bidding history.
Their predicting approach aims to help DSPs fine-tune their bids accordingly.
Though such studies help us understand some internal mechanisms of ADXs and DSPs, they are not applicable to our setting as we try to infer the cumulative ad-cost of each user based on user-related features that are measurable from the user's device over time.
\section{Discussion \& Conclusion}
\label{sec:discussion}

{\bf Limitations.}
Our approach, through \toolname\ plugin, monitors the charge prices for each auctioned ad-slot. 
However, there are several cost models in digital ad-buying.
For example, Cost-Per-Impression is where the advertiser pays when an impression is rendered, and Cost-Per-Click is where the advertiser pays only if the impression is rendered and clicked, etc.
Given that our study is based on passive measurements, we currently unable to determine the cost model of each auctioned ad-slot.
Therefore, we assume all charge prices are under the Cost-Per-Impression model, thus computing the maximum cost advertisers pay for a user.

{\bf Computing The financial worth of individuals.}
Via our methodology, users can estimate, at real time, the cost advertisers pay to reach them.
However, this work's important technical contribution, i.e., \emph{how to compute the financial worth of individuals} with a passive measurement method has several applications. 
Our methodology could provide more transparency on what each type of the users' personal data is worth, and allow users to take advantage of, and (re)negotiate their online value with data hub companies who are interested in investing and innovating in the area of targeted advertising.
Also, such companies can use our methodology to assess the costs implied in this area, how to allocated appropriate resources and, even, estimating bidding strategies of competitors.
In addition, regulators and policy makers could provide guidelines and laws to users and companies for containing the leakage of users' personal data.
Finally, tax auditors could estimate ad-companies' revenues, and detect discrepancies from their tax declarations in an independent and transparent way.\\

\noindent{\bf Conclusion.}
In this study, we aim to enhance transparency in the ad ecosystem, where user's personal data is the most important 
factor affecting the pricing dynamics. We developed a first of its kind methodology to 
estimate how much do advertisers pay to reach a user.
Our methodology leverages the rapidly growing RTB protocol and the new advertising model of programmatic instantaneous auctions, where the advertisers evaluate the users' collected data at real time and bid for an ad-slot in their display.
Our study analyzes the RTB price notifications sent to winning advertising bidders 
and focuses on the distinction between cleartext and encrypted price notifications and how to estimate the latter.
Towards this end, we train a model using as ground truth prices obtained by running our own probing ad-campaigns.
We bootstrap and validate our methodology using a year long trace of real user browsing data, as well as 
two real world  ad-campaigns. 
Finally, we designed \toolname: a system to allow users to compute at real time the value advertisers pay to reach them. 
As future work, we plan to make our prototype (a preliminary version can be seen in Figure~\ref{fig:pluginScreen}) available 
for the community to test and explore its effectiveness with online users.\\

\noindent {\bf \large ACKNOWLEDGEMENTS}\\
The authors would like to acknowledge the contributions and  help  received  during  the  execution  of  this  project:
Prof. Vishal  Misra  and  his  team  (Columbia  Univ.), for a preliminary study as part of an early DTL grant; 
Prof. Evangelos Markatos (FORTH-ICS) and Dr. Claudio Soriente (Telefonica I+D) for their valuable feedback and comments; 
Jose Ramon Gomez Utrilla (Telefonica 4th Platform team), for his help with the design and execution of ad-campaigns;
Costas Iordanou, for his help with the Google AdWords; Rafa Gross-Brown (DTL) on the design of ads;
and Xiaoyuan Yang and  Martin Levi Gonzalez (Telefonica Niji team) for providing the mobile data.

The research leading to these results has received funding from the European Union's Horizon 2020 
research and innovation programme under grant agreements No 653449 (project TYPES) and Marie Sklodowska-Curie 
grant agreement No 690972 (project PROTASIS). The paper reflects only the authors' view and the Agency and the Commission are not responsible 
for any use that may be made of the information it contains.

\balance
\bibliographystyle{ACM-Reference-Format}
\bibliography{cameraRD}

\end{document}